\documentclass[smallextended,final]{svjour3}

\usepackage{amsmath}
\usepackage{amsfonts}
\usepackage{amssymb}

\usepackage[T1]{fontenc} 

\usepackage[english]{babel} 
\usepackage{graphicx} 
\usepackage{subfigure} 
\usepackage{multirow} 
\usepackage{dcolumn} 
\usepackage{float}
\usepackage{afterpage}
\renewcommand{\d}{{\rm d}} 
\newcommand{\td}[2]{\frac{{\rm d} {#1}}{{\rm d} {#2}}} 
\newcommand{\df}{\ {\overset {\rm def} =}\ }

\begin{document}
\title{Formation of Gyrs old black holes in the centers of galaxies within the Lema\^{\i}tre--Tolman model}
\titlerunning{Formation of Gyr old BH in the centers of galaxies within LT model}
\author{Przemys{\l}aw Jacewicz \and Andrzej Krasi{\'n}ski}
\institute{P. Jacewicz \email{pjac@camk.edu.pl}, A. Krasi{\'n}ski \email{akr@camk.edu.pl}
\at Nicolaus Copernicus Astronomical Center, Polish Academy of Sciences, Bartycka 18, 00-716 Warszawa, Poland}
\date{Received: 21 March 2011 / Accepted: 16 August 2011 / Published online: 7 September 2011}
\journalname{General Relativity and Gravitation}
\begin{abstract}
In this article we present a model of formation of a galaxy with a black hole in
the center. It is based on the Lema\^{\i}tre--Tolman solution and is a
refinement of an earlier model. The most important improvement is the choice of
the interior geometry of the black hole allowing for the formation of Gyrs old
black holes. Other refinements are the use of an arbitrary Friedmann model as the background (unperturbed) initial state and the adaptation of the model to an arbitrary density profile of the galaxy.
Our main interest was the M87 galaxy (NGC 4486), which hosts a supermassive
black hole of mass $3.2\cdot 10^{9}M_{\odot}$. It is shown that for this
particular galaxy, within the framework of our model and for the initial state being a perturbation of the $\Lambda$CDM model, the age of the black hole can be up to $12.7$ Gyrs. The
dependence of the model on the chosen parameters at the time of last scattering was also studied. The maximal age of the black hole as a function of the $\Omega_m$ and $\Omega_\Lambda$ parameters for the M87 galaxy can be $3.717$ or $12.708$~Gyr.
\end{abstract}
\PACS{98.80.-k, 
      98.62.Ai, 
      98.62.Js 
      }
\maketitle
\section{Introduction}

Recent years brought increasing evidence that large galaxies host massive black
holes at their centers. The best evidence was obtained for the Milky Way, due to
the proximity of the galactic center located at the distance $R_0 = 8.33\pm
0.35$ kpc. Analysis of orbits of the S-stars cluster near the galactic core,
especially the S2 star, allowed to conclude that a mass of $4.31\pm 0.36\cdot
10^{6}M_\odot$ is responsible for the apparent movement of the stars
\cite{Gillessen:2009a,Gillessen:2009b,Gualandris:2010,Ghez:2008}. Such mass
enclosed in a small volume, of radius less than $0.01$~pc, can only be a black
hole. The existence of black holes at the centers of galaxies calls for
construction of models describing their creation and growth. This work presents
improvements to one of such models \cite{Krasinski:2004a}.

It is thought that the currently existing structures in the Universe (like
galaxies, clusters of galaxies and voids) evolved out of small inhomogeneities
that are observed as directional variations ($\Delta T / T \approx 10^{-5}$) of
the temperature of the cosmic microwave background (CMB) radiation. The
generation of these inhomogeneities is a problem only when one insists on using
models of the Universe that are born spatially homogeneous. In an inhomogeneous
model, such as the Lema\^{\i}tre--Tolman (LT) model used here, the Universe
emerges from the Big Bang being already inhomogeneous. The inhomogeneities are
generated within the Big Bang and their amplitudes are arbitrary parameters that
can be adapted to observational constraints. (But it must be stressed that the
LT model, in which the matter is assumed to be dust, cannot be applied to epochs
earlier than last scattering. A {\em still more general} inhomogeneous model
that includes pressure gradients must be used for the pre-last-scattering
epoch.) It has already been proven in an earlier paper \cite{Krasinski:2004a},
using an LT model, that with a suitable choice of the velocity profile at last
scattering, an object can be created that has the same density profile at
present as the M87 galaxy, and contains a black hole around the center that has
the same mass as the black hole in M87. (For an exposition of the method used in
Ref. \cite{Krasinski:2004a} see Refs. \cite{Krasinski:2001,Krasinski:2004b} and
\cite{Plebanski:2006}.) The black hole is created because matter around the
center expands more slowly than farther away, what leads, at a certain moment,
to the creation of the apparent horizon surrounding a trapped region, and soon
after to the creation of a Big Crunch singularity at the center.

In Ref. \cite{Krasinski:2004a} two particular configurations of the LT model
were considered. In one, the black hole formed around a pre-existing wormhole
within a fraction of a second after the Big Bang. In the other, there was no
wormhole or black hole initially, and it formed during the evolution of the
proto-galaxy. The first configuration is somewhat exotic and we will not pursue
it here. The problem with the second configuration was that the implied age of
the black hole was only $4 \times 10^8$ years, what is inconsistent with
astrophysical implications
\cite{2008MNRAS.391..481S,2006ApJ...643..641H,2004ApJ...614L..25Y}. The reason
for the implied definite age was that the LT model used there contained too few
free parameters.

This paper is an improvement over Ref. \cite{Krasinski:2004a}. The method of
constructing the model is the same as before. We define a certain density or
velocity profile at last scattering, taking care that the amplitude remains
within the limits implied by the measurements of the temperature distribution of
the CMB radiation. We define another density profile at the present time that
agrees with the observationally determined density profile of the M87 galaxy
\cite{Fabricant:1980} and contains a black hole at the center, of mass $3.2\pm
0.9\cdot 10^{9}M_\odot$, equal to the mass of the black hole observed in M87
\cite{Macchetto:1997}. The improvement consists in the LT model having more free
parameters. Thanks to this, the age of the black hole is no longer determined
and can be up to $\sim 12.7$ Gyr, in agreement with what astrophysics tells us
\cite{2008MNRAS.391..481S,2006ApJ...643..641H,2004ApJ...614L..25Y}.\footnote{We
do not wish to enter any discussion of the method by which the age of the black
hole in M87 is inferred. We only wish to demonstrate that, whatever that
inference is, our model can be made consistent with it.}

Our model has certain weaknesses, of which we are aware, but, just as was stated
in Ref. \cite{Krasinski:2004a}, we treat it as an exploratory step into a new
territory: we intend to test a new method that we hope will be improved in the
future. The deficiencies to be removed in the future are the following:

\begin{enumerate}
\item No real galaxy is spherically symmetric. The model can be said to apply
approximately to elliptic galaxies (M87 is one), but spherical symmetry leads to
the next problem listed below.

\item Real galaxies rotate. In a spherically symmetric model rotation is
necessarily zero. The presence of rotation influences the time scale of
evolution, for example, by slowing down the collapse it may significantly delay
the formation of the black hole. However, no exact solutions of Einstein's
equations are known that would describe matter that is expanding and rotating at
the same time. All known expanding solutions have zero rotation, all known
rotating solutions are either stationary or unrealistic for other reasons (for
more on this see Ref. \cite{Plebanski:2006}). So if one wishes to have an
evolving configuration that can be described by the exact formalism of general
relativity, not much is left beyond spherically symmetric models.

\item As was stated in Ref. \cite{Krasinski:2004a}, the perturbation that would
evolve into a single galaxy would have the diameter of approx. 0.004$^{\circ}$
in the CMB sky. The current best angular resolution of measurements of the
temperature fluctuations of the CMB radiation is 0.2$^{\circ}$
\cite{WMAP7:2011a,WMAP7:2011b}. Consequently, the amplitude of fluctuation at
this large angular scale does not give us information that we need to constrain
our model. Lacking any better possibility, just as in the earlier paper
\cite{Krasinski:2004a}, we took care that the amplitudes of our initial density
and velocity profiles do not exceed the limits set at 0.2$^{\circ}$.
\end{enumerate}

The remarks above show that our toy model cannot be literally taken as the
actual model of an existing galaxy. However, it avoids a few other deficiencies
that could be contemplated:

\begin{enumerate}
\item Once the black hole is formed in an LT model, where pressure is zero and
all motions are radial, it keeps accreting matter until it swallows up the whole
mass contained in the region where $E < 0$. This happens in a finite time, which
is arbitrarily long in the neighbourhood of $E = 0$. However, if the function
$E(M)$ has such a profile that at a certain $M = M_g$ $E(M_g) = 0$ and $E(M) >
0$ for $M > M_g$, then the mass from the region $M \geq M_g$ is not accreted
onto the galaxy. Thus, $M_g$, which is independent of time, can be interpreted
as the mass of the galaxy. This is outside the region considered in our paper.
The geometrical radius of the $M = M_g$ surface will be expanding as dictated by
the $E = 0$ evolution equation of the LT model. For the surface of a
galaxy\footnote{Data for the M87 galaxy taken from \cite{Wu:2006}, values of the
constants $c$ and $G$ from http://physics.nist.gov/cuu/Constants/index.html, and
the relations between distance units from
http://www.asknumbers.com/LengthConversion.aspx. All values rounded off.} of
radius 32 kpc $\approx 10^{21}$ m and mass $2.4 \times 10^{12} M_{\odot} = 4.8
\times 10^{42}$ kg, if it were to evolve by the $E = 0$ Lema\^{\i}tre -- Tolman
equation ${R,_t}^2 = 2M/R$, the current velocity of expansion would be approx.
800 km/s $\approx 8.2 \times 10^{-4}$ pc/yr, and constantly decreasing because
$\Lambda = 0$. This is negligible compared to the error in determining the edge
of a galaxy.

\item Each real galaxy is surrounded by vacuum. The cosmological
model takes over at a considerably larger scale. If one wants to illustrate this
situation in a toy model like ours, it is enough to match our LT galaxy model to
the Schwarzschild spacetime at a certain mass $M = M_g$, and then to match the
Schwarzschild region on the outside to another LT or Friedmann region modelling
the Universe. Both matchings would take place outside the region we consider and
would have no influence on what happens inside the galaxy. An explicit example
of such matchings is given in Ref. \cite{Matravers:2001}.
\end{enumerate}

The structure of this paper is as follows. Sec.~\ref{Sec:LT_model} presents the
basic properties of the Lema\^{\i}tre--Tolman model. Sec.~\ref{Sec:basic_model}
briefly outlines the method introduced in
Refs.~\cite{Krasinski:2001,Krasinski:2004b} and emphasises the key elements of
the model. In Sec.~\ref{Sec:development} we describe the improvements to the
basic model: the more general forms of the free functions in the LT model and
the more general (and more realistic) density profile of the galaxy at the
present time. This section also gives the necessary equations for the use of an
arbitrary FLRW model as the background at the initial instant (in Ref. \cite{Krasinski:2004a} the background was assumed to be spatially
flat). Sec.~\ref{Sec:M87} presents the results of application of the model to
the M87 galaxy. Evolution from spatially homogeneous initial profiles of
velocity and density to a galaxy with a black hole is described. A spatially
homogeneous (flat) density profile is obviously within the observational limits
on density perturbations, but may lead to the following problem. After the LT
model is uniquely determined by the initial and final density profiles, the
initial velocity profile that caused the condensation can be {\em calculated}
from the model and may turn out to have a too large amplitude. The same may
happen with the initial velocity profile being flat -- the calculated amplitude
of the initial density profile may turn out to be too large. Graphs shown in
Sec.~\ref{Sec:M87} prove that this did not happen; all implied profiles are
consistent with the observational constraints. Sec.~\ref{Sec:M87} also shows how
the age of the black hole and the arbitrary functions in the LT model change
with the initial FLRW model. The last section contains summary and conclusions.

\section{The Lema\^{\i}tre--Tolman model - basic properties}\label{Sec:LT_model}

The Lema\^{\i}tre--Tolman model is a spherically symmetric nonstatic solution of
the Einstein equations with a dust source
\cite{Lemaitre:1933.1997,Tolman:1934.1997,Bondi:1947}. Its metric is
\begin{equation}\label{Eq:LT_metric}
\d s^2 = \d t^2-\frac{{R,_r}^2}{1+2E(r)}\d r^2-R^2(t,r)(\d\theta^2+\sin^2\theta\d\varphi^2),
\end{equation}
where $E(r)$ is an arbitrary function arising as an integration constant from
the Einstein equations, $R(t,r)$ is a function satisfying
\begin{equation}\label{Eq:LT_evolution_equation}
{R,_t}^2=2E+2\frac{M}{R}+\frac{1}{3}\Lambda R^2,
\end{equation}
where $\Lambda$ is the cosmological constant and $M(r)$ is another arbitrary
function. Equation \eqref{Eq:LT_evolution_equation}, called the evolution
equation, is a first integral of the Einstein equations, $R(t,r)$ is called the
areal radius and $R,_t$ is referred to as the velocity. The matter density
$\rho(t,r)$ is
\begin{equation}\label{Eq:LT_rho}
\frac{8\pi G}{c^4}\rho = \frac{2M,_r}{R^2 R,_r}.
\end{equation}
When $\Lambda=0$, equation \eqref{Eq:LT_evolution_equation} has three
families of solutions depending on the sign of $E(r)$. They are as follows:\\
$E<0$ -- elliptic evolution
\begin{equation}\label{Eqs:LT_elliptic_solution}
\begin{aligned}
R(t,r) &= \frac{M}{-2E}(1-\cos\eta),\\
\eta - \sin\eta &= \frac{(-2E)^{3/2}}{M}(t-t_B(r)),
\end{aligned}
\end{equation}
$E=0$ -- parabolic evolution
\begin{equation}\label{Eq:LT_parabolic_solution}
R(t,r)=\left(\frac{9}{2}M(t-t_B(r))^2\right)^{1/3},
\end{equation}
$E>0$ - hyperbolic evolution
\begin{equation}\label{Eqs:LT_hyperbolic_solution}
\begin{aligned}
R(t,r) &= \frac{M}{2E}(\cosh\eta-1),\\
\sinh\eta - \eta &= \frac{(2E)^{3/2}}{M}(t-t_B(r)),
\end{aligned}
\end{equation}
where $\eta$ is a parameter and $t_B(r)$ is one more arbitrary integration function called the bang time. The
formulas given above are covariant under arbitrary coordinate transformations of
the form $r'=g(r)$, which allows for $r$ to be chosen freely, what in turn means
that one of the three functions $E(r)$, $t_B(r)$ and $M(r)$ can be fixed at our
convenience by the appropriate choice of $g$.

From the form of the solution in the elliptic case,
Eqs.~\eqref{Eqs:LT_elliptic_solution}, we can deduce that $R(t,r)$ reaches a
maximum for each $r$ and then decreases to zero. The values of $\eta$ for which
$R(t,r)$ vanishes are $\eta=0$, corresponding to the Big Bang, and $\eta=2\pi$
corresponding to the Big Crunch at $t=t_C$
\begin{equation}\label{Eq:LT_crunch_time}
t_C(r) = t_B(r) +\frac{2\pi M}{(-2E)^{3/2}}.
\end{equation}
The function $t_C(r)$ is called the crunch time function.

The Friedman--Lema\^{\i}tre--Robertson--Walker (FLRW) models arise from the LT
models in the limit
\begin{equation}\label{Eq:LT_FLRW_limit}
t_B = {\rm const},\qquad E=-\frac{1}{2}kr^2,\qquad M=M_0 r^3,
\end{equation}
where $k$ and $M_0$ are constants.
\subsection{Shell crossings}

It is important to avoid shell crossings, that is loci of $R,_r=0$ and $M,_r\neq
0$, where a constant $r$ shell collides with its neighbour (as a consequence of
$g_{rr}=0$). Shell crossings are curvature singularities and cause the density
to diverge ($\rho\rightarrow\infty$) and change sign to become negative. Such an
undesirable behavior can be avoided if the shapes of the three arbitrary
functions defining the LT model are chosen appropriately. The general conditions
for avoidance of shell crossings are given in \cite{Hellaby:1985,Hellaby:1986}.
\subsection{$M(r)$ as radial coordinate}

It is convenient to use $M(r)$ as the radial coordinate, that is $r'=M(r)$. It is possible because $M(r)$ will be a
strictly growing function, $M,_r>0$, in the whole region under consideration.
Then we have $R=R(t,M)$, $\rho=\rho(t,M)$ and Eq.~\eqref{Eq:LT_rho} becomes
\begin{equation}
\frac{8\pi G}{c^4}\rho = \frac{2}{R^2 R,_M}=\frac{6}{(R^3),_M},
\end{equation}
from which we find
\begin{equation}\label{Eq:LT_areal_radius}
R^3(t,M)-R_i^3 = \frac{3}{4\pi}\int_{M_i}^{M}\frac{\d\xi}{\rho(\xi)},
\end{equation}
where $\rho$ is the density profile of the galaxy as a function of $M$, $R_i$
and $M_i$ are constants to be determined later (see
Sec.~\ref{Sec:recombination_parameters}). Hereafter we use $M$ as the radial coordinate.
\section{A galaxy with a central black hole in the LT model}\label{Sec:basic_model}

An LT model is generally specified by defining the shapes of the arbitrary
functions $E(M)$ and $t_B(M)$, what then allows for the reconstruction of the
evolution of the model by using
Eqs.~\eqref{Eqs:LT_elliptic_solution}--\eqref{Eqs:LT_hyperbolic_solution}.
However, it is possible to follow a different path, that is to calculate these
functions for a specific density or velocity profiles at any two time instants,
as described in \cite{Krasinski:2001} and \cite{Krasinski:2004b}. Moreover,
there are no limitations on those profiles (except the obvious one of spherical
symmetry). The class of the LT evolution applying to each $M$ value can
generally be different, however for the creation of a black hole an LT model
must necessarily be recollapsing in the region near the center of symmetry
$M=0$.

The main idea of the model constructed in \cite{Krasinski:2004a} is that a
spherically symmetric galaxy with a central black hole can be created from
accretion of matter onto an initial fluctuation at recombination. This is the
most probable scenario of structure formation in our Universe
\cite{2003ApJ...582..559V,2002MNRAS.331...98V,2003ApJ...596...47S,2003ApJ...591..499A,2003ApJ...597...21A,2008MNRAS.384....2G}.
Accretion in the model is achieved by a slower expansion near the center, and a
faster one further away. We will now briefly describe the main properties of the
previously developed model.
\subsection{The basic framework of the model}

The basic framework of the model was discussed in \cite{Krasinski:2004a}. The
model allows for the formation of a galaxy with a central black hole by
accretion of background matter onto an initial fluctuation of density or
velocity distribution at the time of last scattering of the CMB. The background
is assumed to be Friedmannian (see Sec.~\ref{Sec:recombination_parameters}). The input data provided for the model are the profiles of the velocity or the
density at last scattering and the current density distribution in the galaxy.
In \cite{Krasinski:2004a} the model was used to create a black hole from a flat
(Friedmannian) velocity distribution at last scattering. In this case, the
fluctuation was due to the corresponding non--flat density profile. The
existence of a black hole at the center of a galaxy is assumed on observational
grounds. The mass of the black hole is an input parameter and likewise must be
taken from observations. The observational density profile of the galaxy was approximated by a simple function, which allowed for
exact calculations.

The construction of the model consisted of the following steps:
\begin{enumerate}
\item determining the functions $E(M)$ and $t_{B}(M)$ in the part outside
the black hole, that is for the mass range $M>M_{\rm BH}$, where $M_{\rm BH}$ is
the observationally determined mass of the black hole in the galaxy,
\item smooth joining of the functions $E(M)$ and $t_{B}(M)$ in the parts
inside and outside the black hole i.e. at the boundary $M=M_{\rm BH}$. The part
inside the black hole is arbitrary and was assumed to be a simple $E<0$
recollapsing model defined by a choice of the bang and crunch time functions,
\item evolution reconstruction - using the full $E(M)$ and $t_{B}(M)$ functions
the evolution of $R(t,M)$ and $\rho(t,M)$ was reconstructed using
Eqs.~\eqref{Eqs:LT_elliptic_solution}--\eqref{Eqs:LT_hyperbolic_solution}.
\end{enumerate}

In the model the time of the black hole formation is the value of the future
apparent horizon at its minimum i.e. $(t_{\rm AH+})_{\rm min}$. However, in this
model the crunch time function and the future apparent horizon almost coincide,
the instant of the creation of the black hole is very close to $t_C(0)$ (see
Fig.~\ref{Fig:bhspace2} and \cite{Krasinski:2004a} for details). The first
application of the model was to the M87 galaxy which hosts a supermassive black
hole of mass $3.2\cdot 10^{9}M_{\odot}$ \cite{Macchetto:1997}, with density
profile being an approximation of Fabricant et. al. \cite{Fabricant:1980}. In
the first parametrization of the black hole's interior geometry the value of
$(t_{\rm AH+})_{\rm min}$ was high, leading to a black hole of the age of a few
hundred million years, which is in disagreement with other estimations of
supermassive black hole's age based on the $\Lambda$CDM model
\cite{2008MNRAS.391..481S,2006ApJ...643..641H,2004ApJ...614L..25Y}.
\section{Development of the model}\label{Sec:development}

The model allows for modifications which can lead to more realistic predictions.
Particular attention was paid to arrive at such bang and crunch time functions
that would give the black hole's age of the order of Gyr.
\subsection{The black hole interior}

\begin{table}[b]
\begin{tabular}{c|c|c}
Name & $t(M)$ & $p_i$ \\
\hline \hline
exp & $p_1 + p_2 \exp\left(p_3 M + p_4\right)$ &
\begin{tabular}{c}
$p_1\sim t(M_{\rm BH})$\\
$p_2\sim 10^0$\\
\end{tabular}\\
hyp & $p_1 + p_2 \cosh\left(p_3 M + p_4\right)$ &
\begin{tabular}{c}
$p_3\sim 10^2$\\
$p_4\sim 10^2$\\
\end{tabular}\\
\end{tabular}
\caption{Parametrizations of the interior bang and crunch time functions. The
table shows the bang and crunch time functions inside the black hole. The third
column gives a general order of magnitude of the value of each parameter.
$t(M_{\rm BH})$ means that the value of the parameter is of the order of the
value of the bang or crunch time function at $M=M_{\rm BH}$. For $t_B(M)$ the
parameters $p_i$ are denoted $b_i$, and for $t_C(M)$ by $c_i$.}
\label{Tab:BH_interior}
\end{table}

The first modification concentrates on the functions $t_B(M)$ and $t_C(M)$ that
define the black hole's interior, for which, for fundamental reasons, no
observational data exist. Therefore, we are free to choose any parametrization
and the only condition is to join it smoothly to the galaxy model. In the most
probable scenarios of black hole creation, it is formed by a collapse of a
massive object. We used exponential and hyperbolic functions to parameterize the
interior functions $t_B(M)$ and $t_C(M)$, see Table~\ref{Tab:BH_interior}. The
chosen forms of the functions have $8$ parameters: $b_1$, $b_2$, $b_3$, $b_4$
for $t_B(M)$ and $c_1$, $c_2$, $c_3$, $c_4$ for $t_C(M)$.

A smooth match of the interior and exterior requires the continuity of the
functions $E(M)$ and $t_B(M)$ and their derivatives at the boundary $M=M_{\rm
BH}$ (the Darmois/Lichnerowicz junction conditions), this is also a sufficient
condition \cite{Bonnor:1981}. Therefore, the following system of equations has
to be solved for the parameters $b_i$ of $t_B(M)$ and $c_i$ of $t_C(M)$,
$i=1,\ldots,4$:
\begin{equation}\label{Eqs:matching_system}
\begin{aligned}
t_B(M_{\rm BH},b_1, b_2, b_3, b_4) &=\left[t_B\right]_{M=M_{\rm BH}},\\
\td{t_B}{M}\Bigl( M_{\rm BH},b_1, b_2, b_3, b_4\Bigr) &=
\left[\td{t_B}{M}\right]_{M=M_{\rm BH}},\\
E(M_{\rm BH},b_1,b_2,b_3,b_4, &\\
c_1,c_2,c_3,c_4) &= \left[E\right]_{M=M_{\rm BH}},\\
\td{E}{M}\Bigl(M_{\rm BH},b_1,b_2,b_3,b_4, &\\
c_1,c_2,c_3,c_4\Bigr) &= \left[\td{E}{M}\right]_{M=M_{\rm BH}},
\end{aligned}
\end{equation}
where the expressions on the left hand side are the analytical functions
depending on the chosen parametrization and the expressions on the right hand
side are the numerically found values of $t_B(M)$ and $E(M)$ for $M=M_{\rm BH}$,
using the method described in \cite{Krasinski:2004a}. The analytical expression
for $E(M)$ is based on Eq.~\eqref{Eq:LT_crunch_time}. However, the assumed bang
and crunch time functions have eight unknown parameters, whereas from
\eqref{Eqs:matching_system} only four parameters can be determined. Therefore,
the values of the other four parameters have to be supplied. To accomplish that
we choose any two parameters for $t_B(M)$ and any two parameters for $t_C(M)$,
and set their values using random numbers. The difficulty here (from the
numerical point of view) is to determine the order of magnitude of the
parameters, but this can be achieved by examination of the type of parameterized
equation (see column $3$ of Table~\ref{Tab:BH_interior}). The system of
equations \eqref{Eqs:matching_system} has to be solved numerically. The usage of
random numbers does not have any impact on the analysis as the multidimensional
minimalization methods, employed further for finding extreme LT models take the
random values only as the starting values for subsequent optimalization.

\subsection{The density profile of the galaxy}\label{Sec:galaxy_density_profile}

A galaxy in the constructed model is described by a density profile and the mass
of the black hole. However, the quantity used in the numerical calculations is the areal radius $R(t_0, M)$ corresponding to that (spherically symmetric)
density profile, where $t_0$ is the present age of the Universe. The equation
used for finding its value is \eqref{Eq:LT_areal_radius}, with $\rho$ being a
function of $M$. Therefore, the general $r$ coordinate representing the distance
from the center of symmetry must be replaced by the mass $M$ of the sphere of
radius $r$. In \cite{Krasinski:2004a} the relation $r=r(M)$ was found
analytically, using an analytically convenient approximation to the
observational density profile of the galaxy. However, we would like to apply the
model to any galaxy with known density profile which is suspected of hosting a
black hole of known mass. In such a general situation, galaxies are described by
density profiles which do not allow for analytical calculation of the relation
$r=r(M)$. Therefore, we must resort to numerical calculations and take $\rho(M) \df \rho(r(M))$, where $r(M)$ is the inverse function to $M(r)$ and is found numerically.

In order to find the areal radius $R(t_0, M)$, the constants $M_i$ and $R_i$ of
Eq.~\eqref{Eq:LT_areal_radius} have to be appropriately chosen. We may put
$R_i=0$, thus assuming $R(t_0,M_i)=0$. Further we specify $M_i$ to be $M_S$ --
the mass already swallowed by the singularity. On a spacetime diagram in the
coordinates $(M,t)$ that mass is represented by the value of $M$, for which the
line $t=t_0$, representing the present moment, intersects the crunch time
function, that is $t_0=t_C(M_S)$ (see Fig.~\ref{Fig:bhspace2}). Therefore
Eq.~\eqref{Eq:LT_areal_radius} becomes
\begin{equation}
R^3\left(t_0, M\right)=\frac{3}{4\pi}\int_{M_{S}}^M\frac{\d\xi}{\rho\left(\xi\right)}.
\end{equation}
The value of $M_S$ can be calculated using the properties of the apparent
horizon (see \cite{Krasinski:2004a}, \cite{Plebanski:2006}). Only for $\rho(M)$
given in a fairly simple form analytical calculations are possible. Here, we do
not find the value of $M_S$ and make use of the properties of the apparent
horizon.

The apparent horizon consists of the events for which
\begin{equation}
R(M_{\rm BH})=2M_{\rm BH},
\end{equation}
where $M_{\rm BH}$ is the mass of the black hole
\cite{Plebanski:2006,Szekeres:1975}. We also have $M_S<M_{\rm BH}$, that is the
mass swallowed up by the singularity must necessarily be smaller than the mass of the black hole. Therefore we have
\begin{equation}
R^3(t_0,M)=\frac{3}{4\pi}\int_{M_S}^{M_{\rm
BH}}\frac{\d\xi}{\rho\left(\xi\right)}+\frac{3}{4\pi}\int_{M_{\rm
BH}}^M\frac{\d\xi}{\rho\left(\xi\right)}.
\end{equation}
The first term, from the definition of the apparent horizon, is equal to
$(2M_{\rm BH})^3$ and the equation for calculating the value of areal radius for
the galaxy under consideration becomes
\begin{equation}\label{Eq:R_now_galaxy}
R(t_0,M)= \left[(2M_{\rm BH})^3+\frac{3}{4\pi}\int_{M_{\rm BH}}^M\frac{\d\xi}{\rho\left(\xi\right)}\right]^{1/3}.
\end{equation}
The advantage of this equation is that it can be applied to any galaxy with known density profile and known mass of the black hole.
\subsection{The parameters at the recombination}\label{Sec:recombination_parameters}

\begin{figure}
\subfigure[]{\includegraphics[width=0.5\columnwidth]{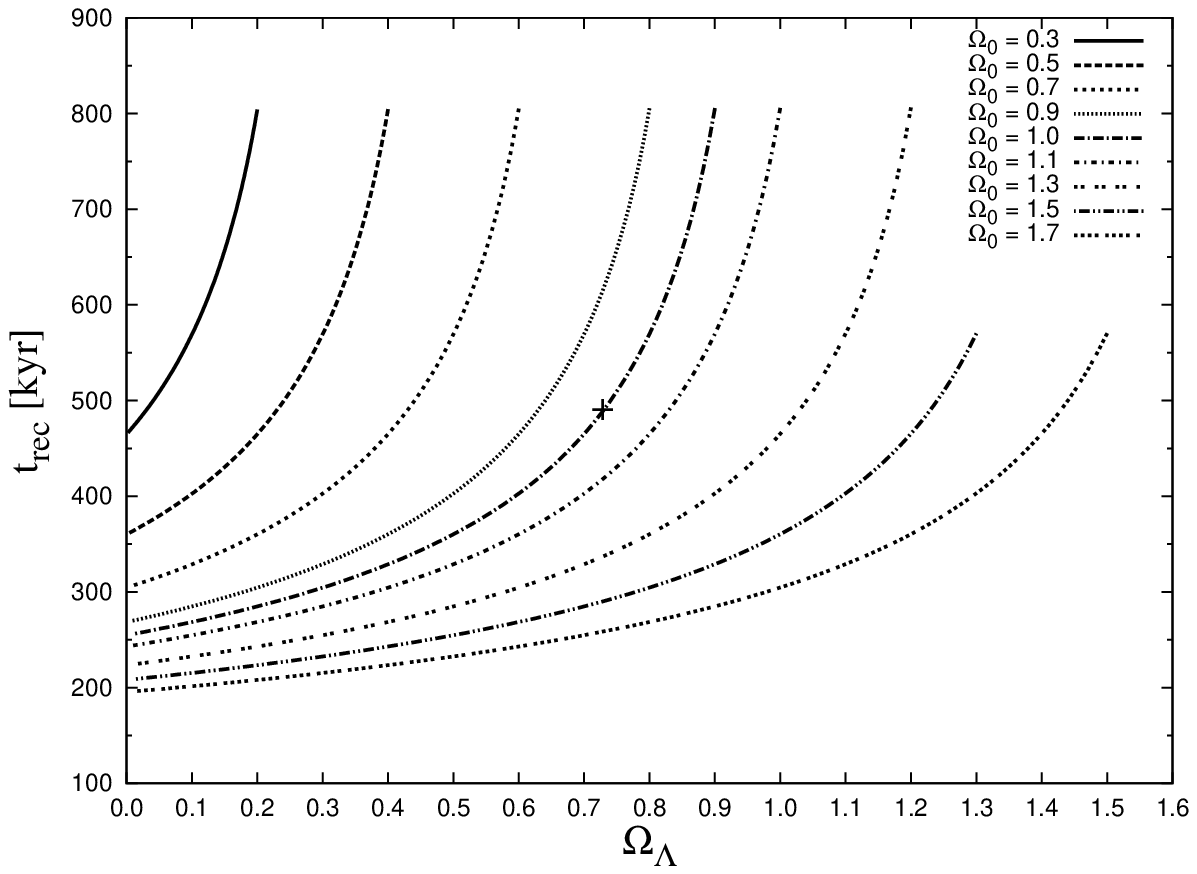}\label{Fig:FLRW_model_plots-a}}
\subfigure[]{\includegraphics[width=0.5\columnwidth]{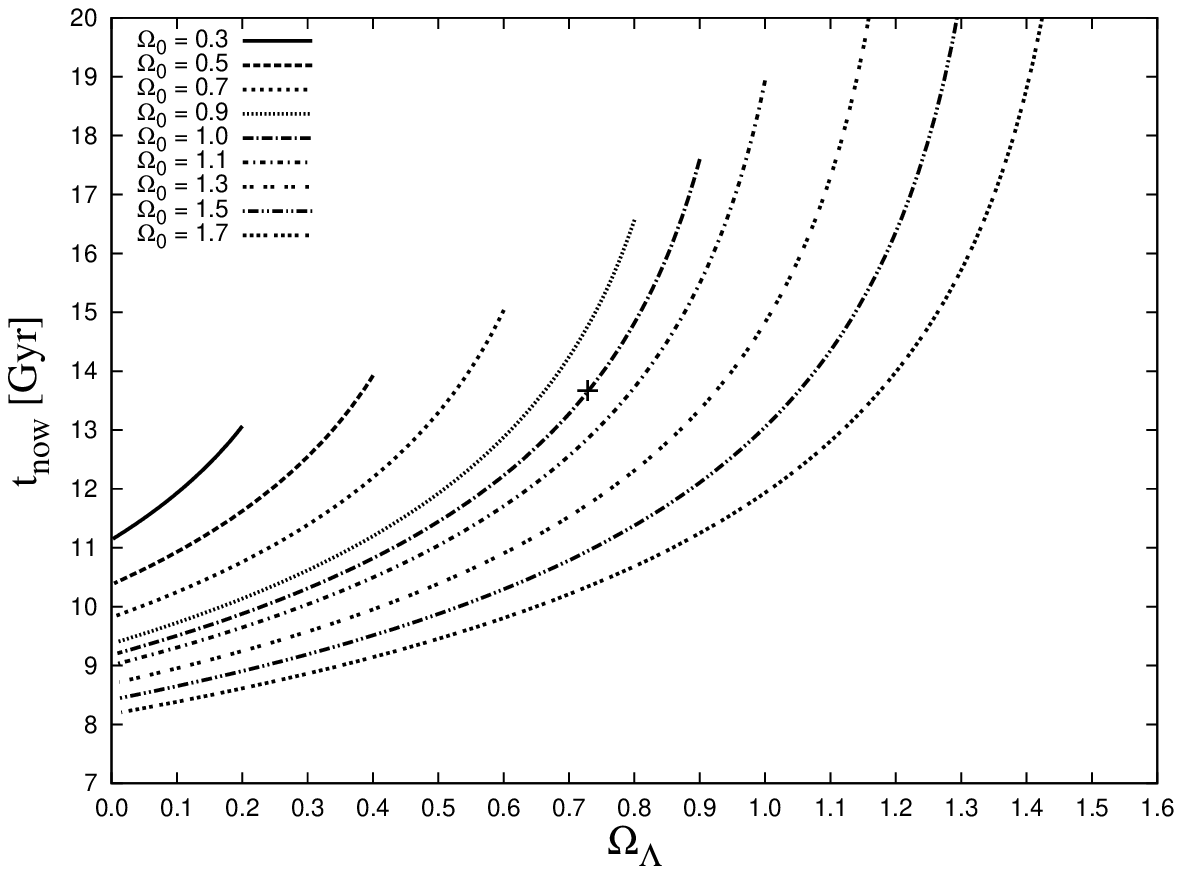}\label{Fig:FLRW_model_plots-b}}
\subfigure[]{\includegraphics[width=0.5\columnwidth]{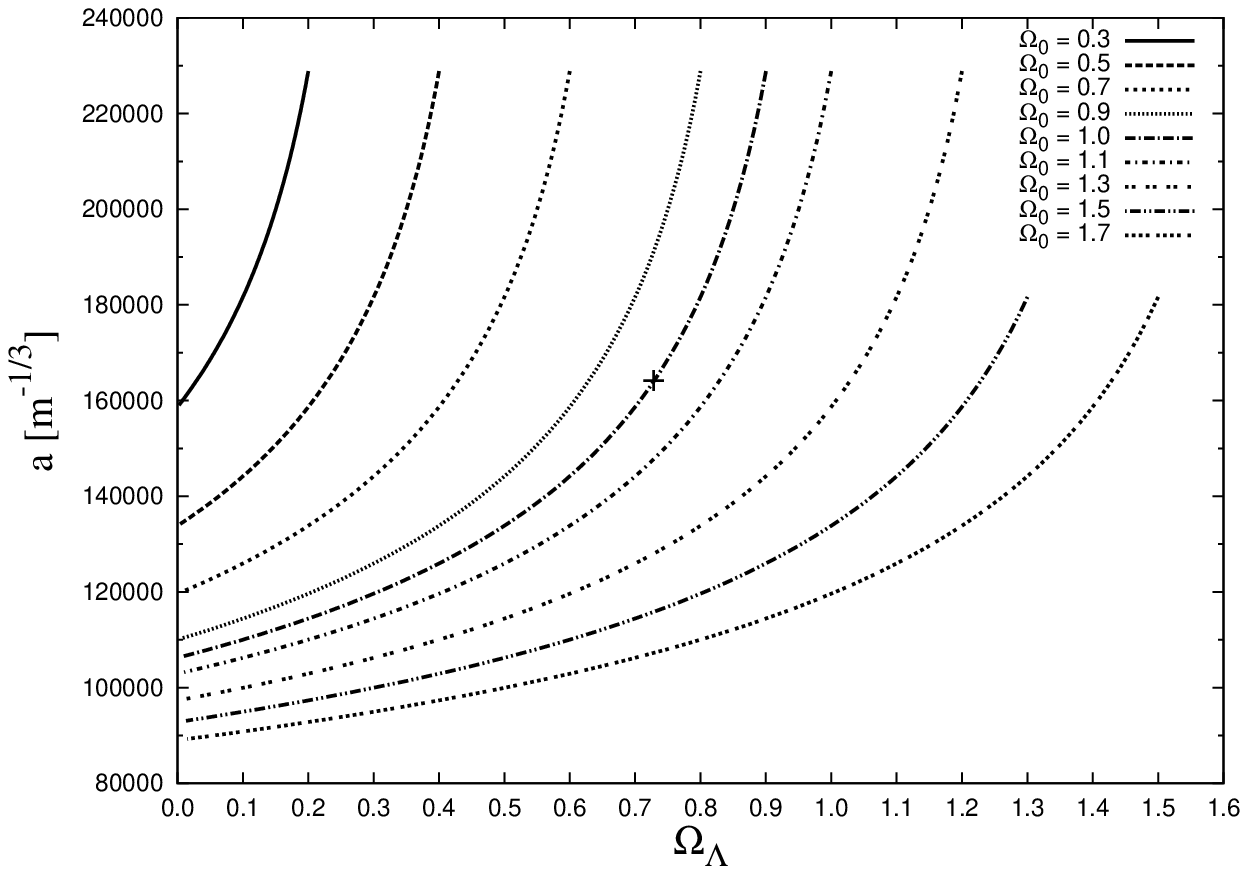}\label{Fig:FLRW_model_plots-c}}
\subfigure[]{\includegraphics[width=0.5\columnwidth]{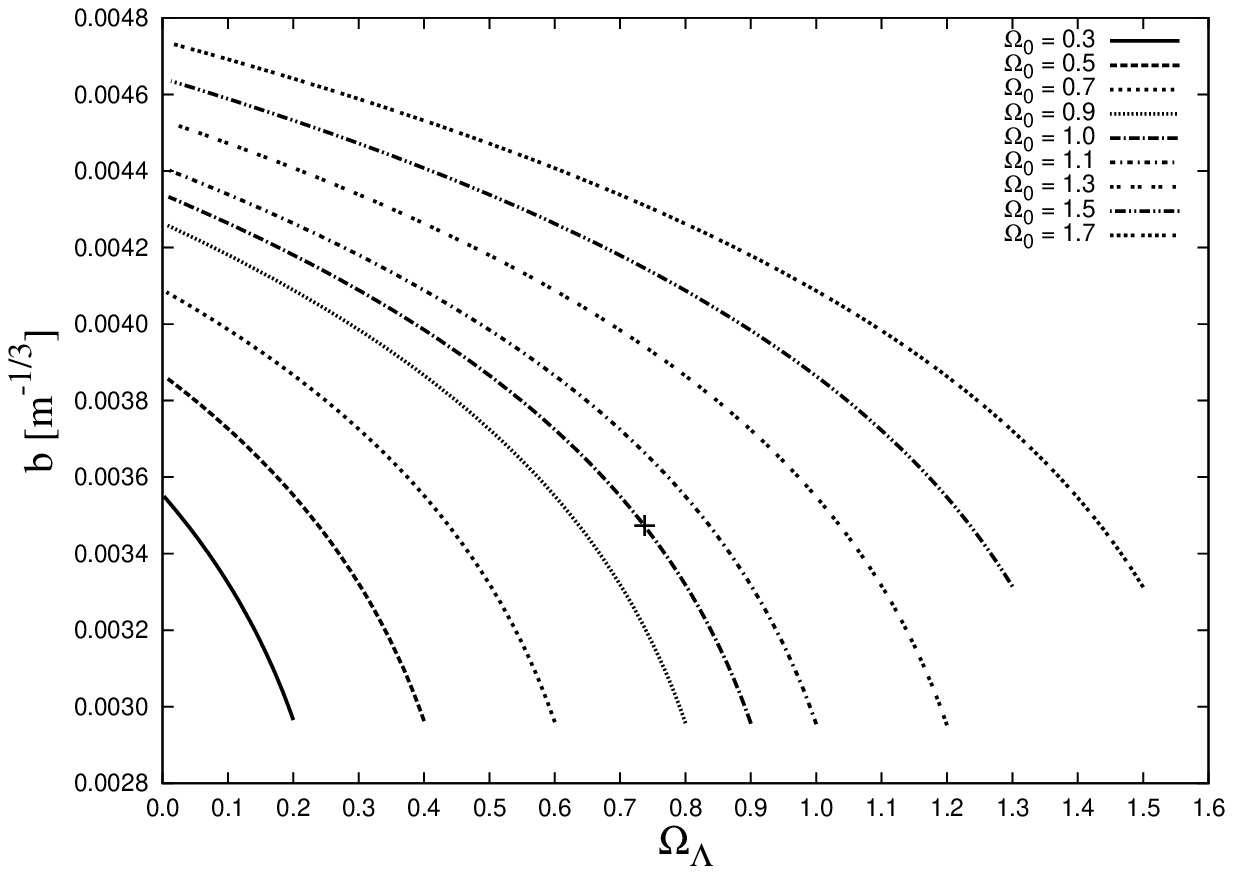}\label{Fig:FLRW_model_plots-d}}
\caption{\subref{Fig:FLRW_model_plots-a} The time of last scattering of CMB in
the FLRW model as a function of $\Omega_\Lambda$ for different values of
$\Omega_0$. For the $\Lambda$CDM model with WMAP 7-year results the instant of
last scattering is at $t_{\rm rec}=494$~kyr, which is marked with a cross in the
plot. \subref{Fig:FLRW_model_plots-b} The age of the Universe for the same set
of $\Omega_\Lambda$ and $\Omega_0$ values. The cross at $t_{\rm now}=13.727$~Gyr
indicates the age of the Universe for the $\Lambda$CDM model with WMAP 7-year
results. \subref{Fig:FLRW_model_plots-c} $a=R/M^{1/3}$ and
\subref{Fig:FLRW_model_plots-d} $b=R,_t/M^{1/3}$ at the time of last scattering
of CMB in the FLRW model for the same set of $\Omega_\Lambda$ and $\Omega_0$
values. For the $\Lambda$CDM model with WMAP 7-year results $a$ equals
$1.649\cdot 10^{5}$~m$^{-1/3}$, while $b$ is equal to $3.483\cdot
10^{-3}$~m$^{-1/3}$.}\label{FLRW_model_plots}
\end{figure}

In \cite{Krasinski:2004a} the parameters at the time of last scattering where
chosen to match those of the $k=0$ FLRW model. In this work we will use general FLRW models with $k\neq 0$ and with
nonvanishing cosmological constant. The areal radius $R(t,M)$ and the scale
factor $S(t)$ of the FLRW model are connected by
\begin{equation}
R(t,M) = r(M)\,S(t),
\end{equation}
with $M$ as the radial coordinate. Using the limiting conditions for the FLRW model arising from
the LT model, i.e. Eq.~\eqref{Eq:LT_FLRW_limit}, we have
\begin{equation}
R(t,M) = M_0^{-1/3}M^{1/3}S(t),
\end{equation}
where the constant $M_0$ determines the particular FLRW model. The value of this
constant can be found by comparison of the solutions of the LT model
(Eqs.~\eqref{Eqs:LT_elliptic_solution}--\eqref{Eqs:LT_hyperbolic_solution}) with
the solutions of the FLRW model \cite{Plebanski:2006}, obtaining
\begin{equation}
M_0 = \frac{4}{3}\frac{G}{c^2}\pi\rho(t) S^3(t),
\end{equation}
where $\rho(t)$ is the (homogeneous) density. By taking the values of $\rho(t)$
and $S(t)$ at the present moment $t=t_0$, that is
\begin{equation}
\rho(t_0)= \Omega_m\,\rho_{\rm crit} = \Omega_m\,\frac{3H_0^2}{8\pi G},\qquad S(t_0)=1,
\end{equation}
where $\Omega_m$ is the density parameter of matter (with equation of state
$p=0$), $\rho_{\rm crit}$ is the critical density with the current estimate of
the Hubble constant $H_0=71\pm 2.5$~km~s$^{-1}$~Mpc$^{-1}$, based on the latest
WMAP results \cite{WMAP7:2011a}, we get the value for $M_0$ as
\begin{equation}
M_0 = \frac{1}{2c^2}\Omega_m H_0^2.
\end{equation}

The value of the scale factor $S(t)$ at the time of recombination (last
scattering of CMB) is found from
\begin{equation}\label{Eq:rec_scale_factor}
S(t_{\rm rec}) = \frac{S(t_0)}{1+z_{\rm rec}}=\frac{1}{1+z_{\rm rec}},
\end{equation}
where $z_{\rm rec}=1088.2\pm 1.2$ is the redshift of CMB from the WMAP results.
The derivative of the scale factor is determined by the Friedmann equation
\begin{equation}\label{Eq:Friedmann}
\left(\frac{{S,_t}}{S}\right)^2 = \frac{8\pi G}{3}\rho_m - \frac{k}{S^2} + \Lambda,
\end{equation}
with the constant $k$ given by
\begin{equation}
k = H_0^2 S^2(t_0)(\Omega_m + \Omega_{\Lambda} - 1) = H_0^2(\Omega_0 - 1),
\end{equation}
where $\Omega_\Lambda=(3/8\pi G)(\Lambda/\rho_{\rm crit})$ and
$\Omega_0=\Omega_m+\Omega_\Lambda$.

We would like to find the dependence of our model on the $\Omega_m$ and $\Omega_\Lambda$ parameters of the FLRW model used to determine the functions at the time of last scattering.
In order to do that we need to know the age of the Universe and the
recombination time for FLRW models with different $\Omega_m$ and
$\Omega_\Lambda$. From the Friedmann equation, using the normalization
$S(t_0)=1$, the age of the Universe $t_0$ is determined by
\begin{equation}\label{Eq:FLRW_t_now}
t_0 = H_0^{-1}\int_0^{1}\frac{\d\xi}{\sqrt{\Omega_m/\xi -
(\Omega_m+\Omega_\Lambda-1)+\Omega_\Lambda\xi^2}}.
\end{equation}
The instant of last scattering of CMB, used as one of the input parameters for
the model, is dependent on the matter content of the Universe. This instant is
defined by such density of the radiation that the energy of the photon becomes
smaller than the ionization energy of the hydrogen atom. As the radiation and
matter have different equations of state, the instant of last scattering is
different for different density parameters. The time of last scattering of CMB
can be found from the equation similar to Eq.~\eqref{Eq:FLRW_t_now}, but with
the upper limit in the integral replaced by $S(t_{\rm rec})$, taken from
Eq.~\eqref{Eq:rec_scale_factor}. Figure~\ref{FLRW_model_plots} shows the age of
the Universe and the instant of recombination as a function of $\Omega_\Lambda$
for different values of $\Omega_0$. This figure also shows the dependence of the
two functions used as input data for the model $a(M)=R(M)/M^{1/3}$ and
$b(M)=R,_t(M)/M^{1/3}$ at the recombination on the value of $\Omega_\Lambda$ for
the same set of $\Omega_0$ values. The two parameters are constant within a FLRW
model and their values for the $\Lambda$CDM model, in units described in
\ref{Sec:Units} and with WMAP 7-year results, are $a=1.649\cdot
10^{5}$~m$^{-1/3}$ and $b=3.483\cdot 10^{-3}$~m$^{-1/3}$.
\subsection{The units}\label{Sec:Units}

Throughout this article we have used the geometric units where both $c$ and $G$
are equal to unity. In order to avoid large numerical values we used
$10^{11}M_\odot$ as the unit of mass. This value corresponds roughly to the mass
of the M87 galaxy.
\section{Application of the model to the M87 galaxy}\label{Sec:M87}

We have tested the constructed model on the M87 (NGC 4486) galaxy, which hosts a
supermassive black hole of mass $3.2\cdot 10^{9}M_{\odot}$ and is about $16$~Mpc
away from the Earth \cite{Macchetto:1997}. The galaxy is mainly known for its
highly relativistic jet, but we are only interested in its density profile and
the mass of the supermassive black hole in the center.
\subsection{The M87 input parameters}

\begin{figure}
\subfigure[]{\includegraphics[width=0.5\columnwidth]{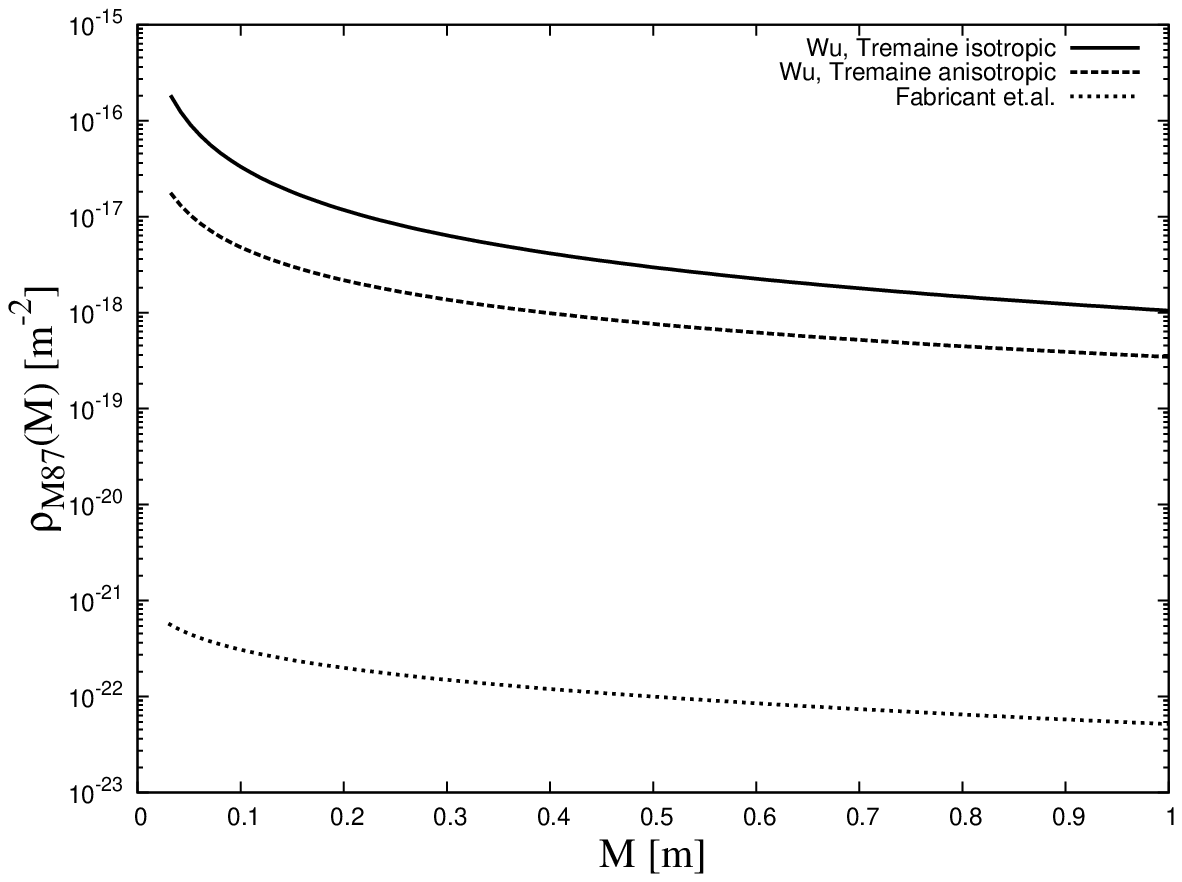}\label{Fig:M87_galaxy_plots-a}}
\subfigure[]{\includegraphics[width=0.5\columnwidth]{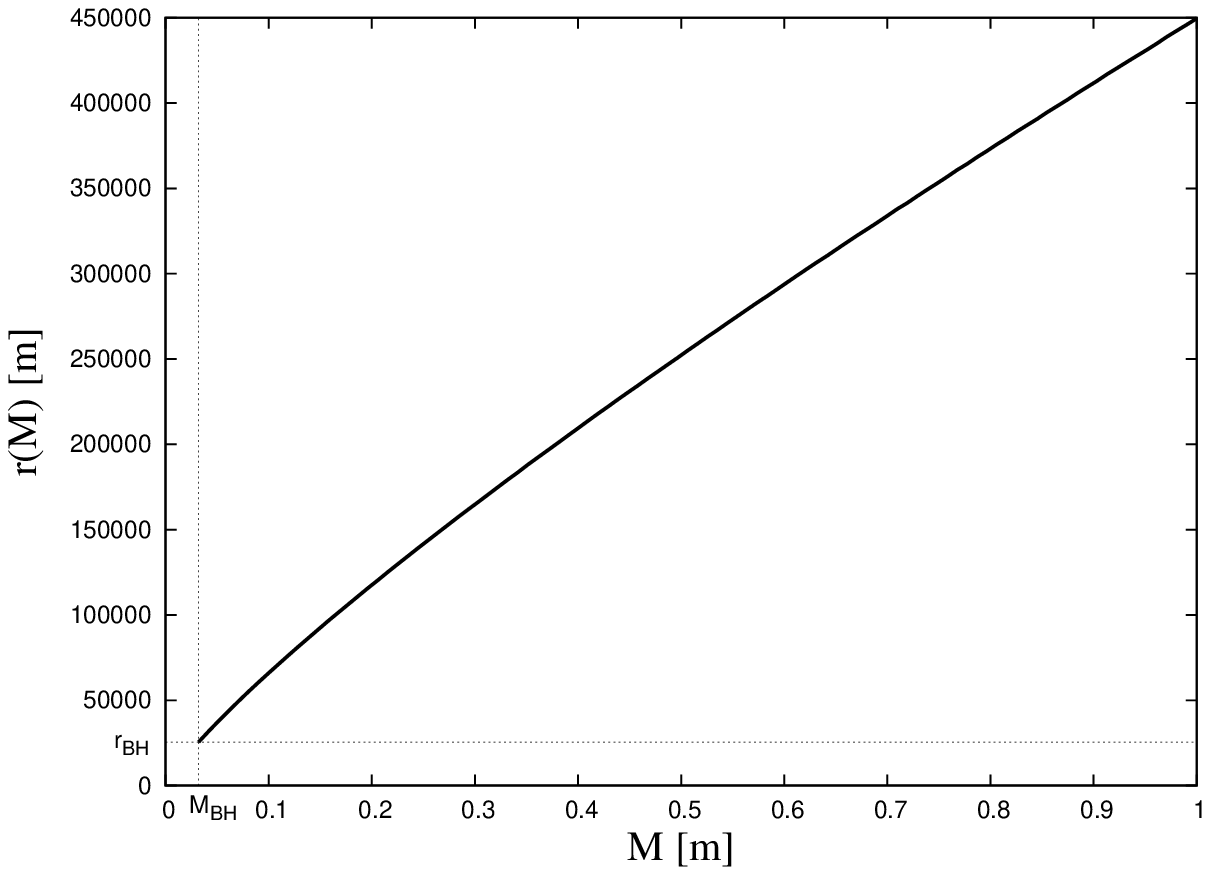}\label{Fig:M87_galaxy_plots-b}}
\subfigure[]{\includegraphics[width=0.5\columnwidth]{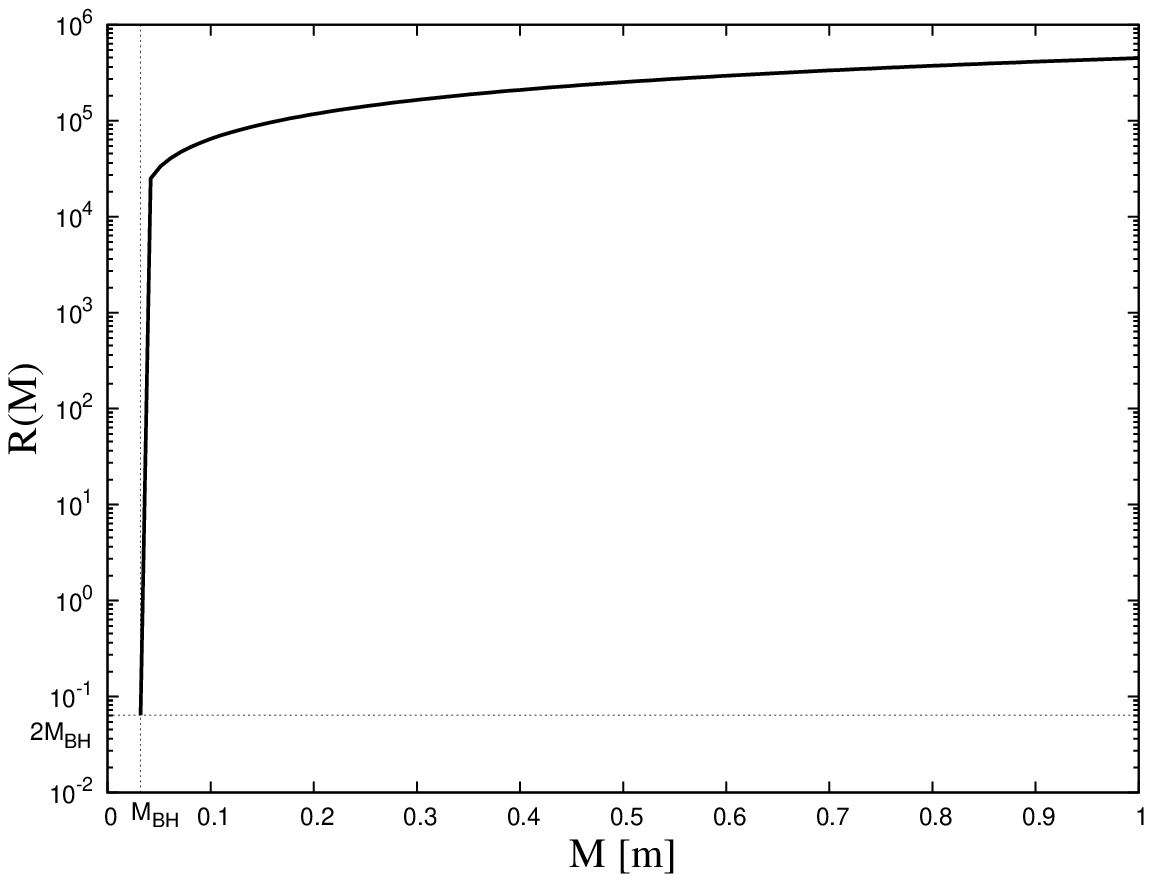}\label{Fig:M87_galaxy_plots-c}}
\subfigure[]{\includegraphics[width=0.5\columnwidth]{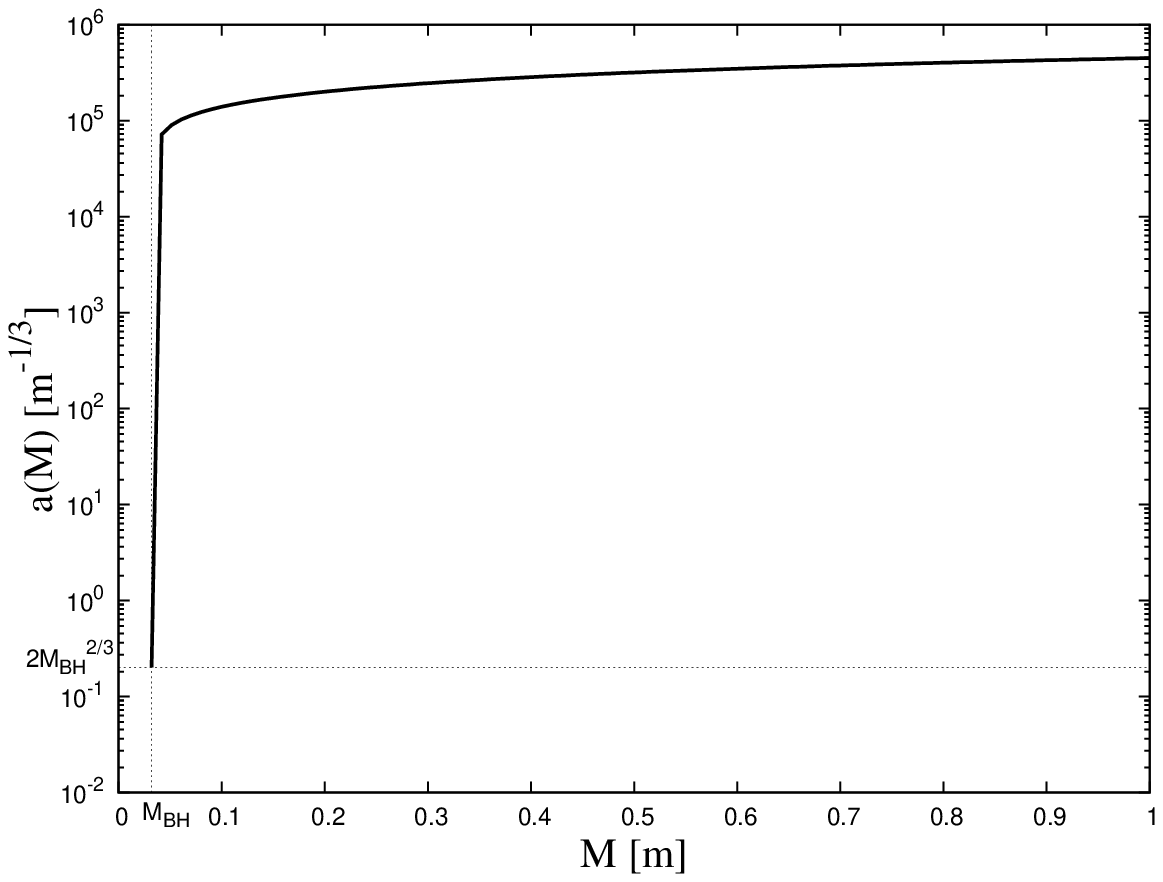}\label{Fig:M87_galaxy_plots-d}}
\caption{\subref{Fig:M87_galaxy_plots-a} Comparison of M87 density profiles. All
curves start at $M=M_{\rm BH}$, which for Fabricant~et.~al. is equal to
$0.03$~m, and $0.032$~m for both Wu,~Tremaine profiles. The profile used in our
calculation is 'Wu, Tremaine isotropic'. Both axes are in geometric units.
\subref{Fig:M87_galaxy_plots-b} The distance--mass relation for the M87 galaxy
with the 'Wu,~Tremaine~isotropic' density profile. The mass equal to the mass of
the black hole is enclosed within the radius $r_{\rm BH}=25538.6$~m.
\subref{Fig:M87_galaxy_plots-c} $R(M)$ and \subref{Fig:M87_galaxy_plots-d}
$a(M)$ functions for the chosen density profile of
Eq.~\eqref{Eq:M87_wu_tremaine_isotropic_profile}. The dotted vertical line
represents the black hole's surface at $M_{\rm
BH}=0.032$~m.}\label{Fig:M87_galaxy_plots}
\end{figure}

In \cite{Krasinski:2004a} the authors used the approximated profile of
Fabricant~et.~al.~\cite{Fabricant:1980}, with the mass of the black hole $M_{\rm
BH}=0.03$~m. As was described in Sec.~\ref{Sec:galaxy_density_profile}, we no
longer have to use any approximation to the density profile and generally we
could use the original, not approximated profile, and the value of $M_{\rm BH}$
for our numerical calculations. However, in recent years the standard value of
$M_{\rm BH}$ and the shape of the density profile for M87 have changed (see
Fig.~\ref{Fig:M87_galaxy_plots-a}). We will use the updated values.

The M87 galaxy is a radio source, and has been the subject of many measurements based on different
methods. In this work we used the 'isotropic' profile of
Wu,~Tremaine~\cite{Wu:2006} based on globular clusters data, given by
\begin{equation}\label{Eq:M87_wu_tremaine_isotropic_profile}
\rho(r) = \rho_0 \left(\frac{r}{r_0}\right)^{-\alpha},
\end{equation}
with the values of the parameters: $r_0=19$~kpc, $\rho_0=1.9\cdot 10^7
M_\odot$~kpc$^{-3}$ and $\alpha=1.8$. The mass of the black hole is assumed to
be $3.2\cdot 10^9 M_\odot$. The 'anisotropic' profile based on the same data,
but modified method, differs only in the value of $\alpha$, which for this
profile is equal to $1.6$. The three density profiles are plotted in
Fig.~\ref{Fig:M87_galaxy_plots}.  Note the change in the magnitude of the
density values.

The quantities used in our calculations that describe the M87 galaxy, besides $M_{\rm BH}$, are the distance--mass relation $r(M)$,
the areal radius $R(t_0,M)$ and $a(M)=R/M^{1/3}$. The distance--mass relation is
found numerically by integration of the density profile and the areal radius
$R(t_0,M)$ is found using Eq.~\eqref{Eq:R_now_galaxy}. The three functions are
plotted in Figs.~\ref{Fig:M87_galaxy_plots-b}--\ref{Fig:M87_galaxy_plots-d}.
\subsection{The scheme of the method}

The general scheme of our method is as follows. Firstly, for the given velocity
or density profile at the last scattering and the M87 density profile (together
with the assumed $M_{\rm BH}$), the functions $E(M)$ and $t_B(M)$ (and also the
resulting $t_C(M)$) are found in the part outside the black hole, that is for
$M>M_{\rm BH}$. Secondly, for the chosen type of black hole interior, that is
the bang and crunch time functions for $M<M_{\rm BH}$ described by the equations
in Table~\ref{Tab:BH_interior}, random four parameters of those equations are
chosen and their values are set using uniformly distributed random numbers in
the appropriate range, depending on the parameter. Then, the values of the other
four parameters are found by numerically solving the equations
\eqref{Eqs:matching_system}. This is done using multidimensional root finding
method based on the Hybrid Method. In this way an LT model model near $M=0$ is
found (necessarily recollapsing), which will serve as the starting point for
further actions. With some density profiles (at $t={\rm now}$ and at last
scattering) it may turn out that the model is not recollapsing at present. This
step is necessary to make sure that this does not happen.

As stated before, for galaxies that host supermassive black holes, such as M87,
we are mainly interested in finding such an LT model, that would yield a black
hole of the age of Gyr. Therefore the age of the black hole is the parameter to
be optimized. Optimization is done using the Simplex algorithm of Nelder and
Mead for minimization of multidimensional functions \cite{Nelder:1965} by again
choosing arbitrary four parameters of bang and crunch time functions in the
interior of the black hole (found in the previous step) and finding such values
of those parameters that would give a black hole of maximal age. For each run we
have also found the LT model of minimal age of the black hole, just as a
comparison to the maximal one. After this we have performed the reconstruction
of evolution together with the calculation of other valuable data.
\subsection{The $\Lambda$CDM model of initial fluctuation}

\begin{figure}
\subfigure[]{\includegraphics[width=0.5\columnwidth]{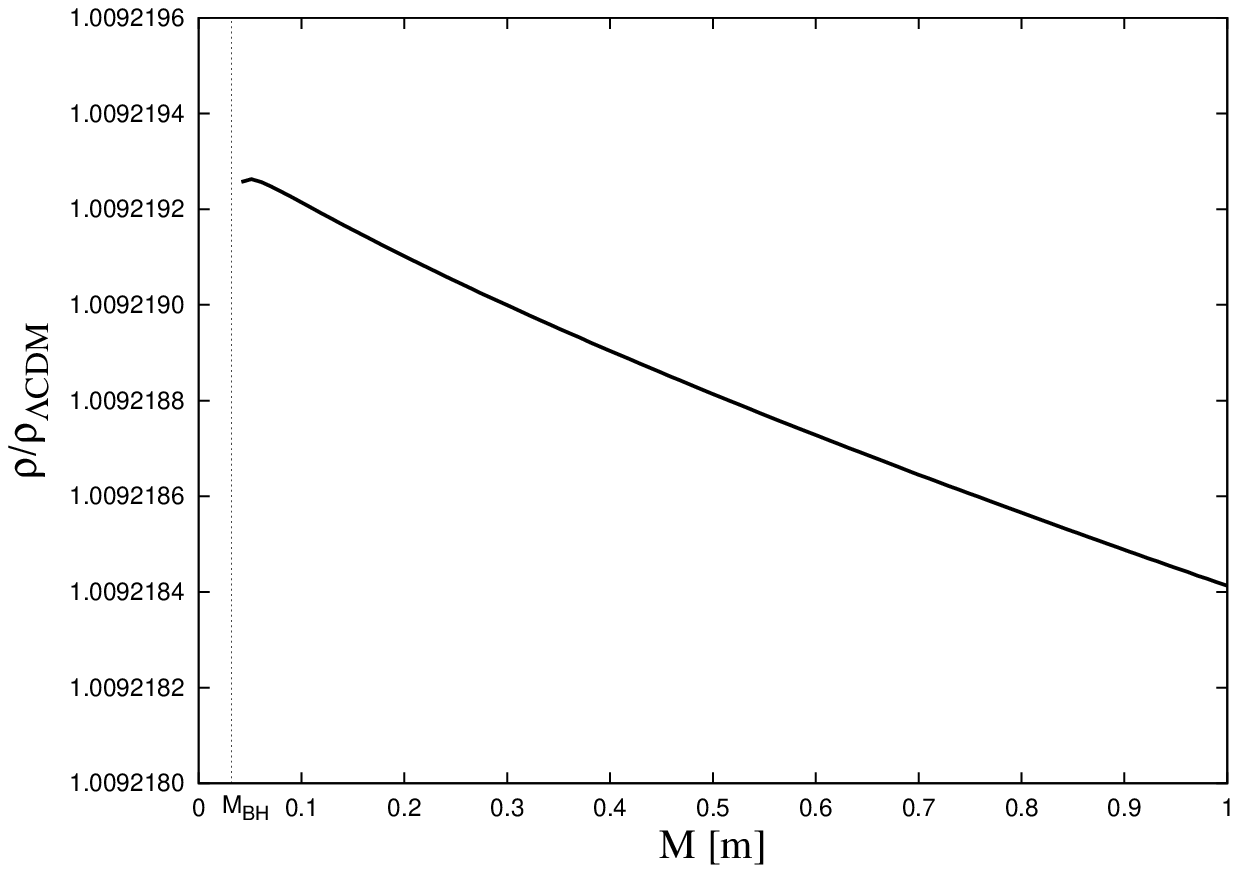}\label{Fig:rho_M_perturbation_velocity_M87_LCDM}}
\subfigure[]{\includegraphics[width=0.5\columnwidth]{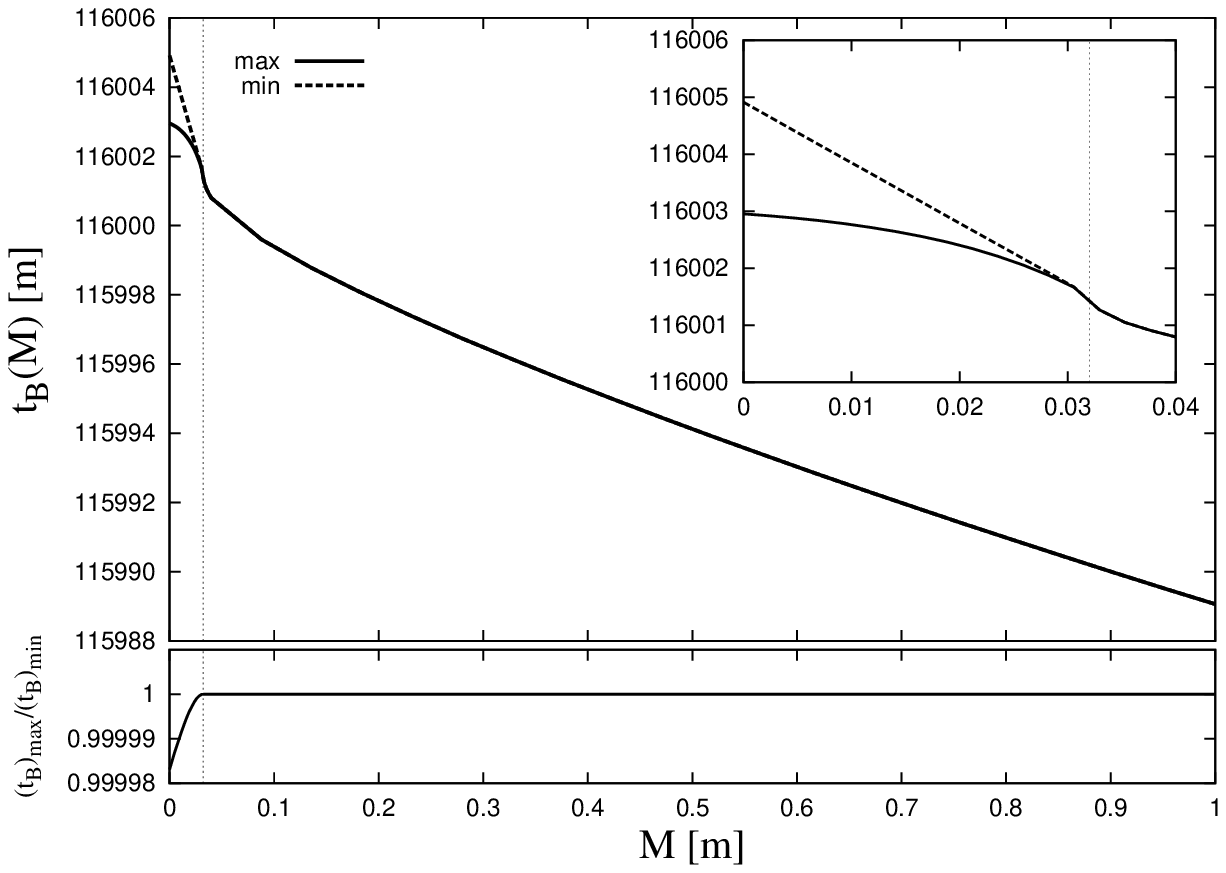}\label{Fig:t_B_velocity_M87_LCDM}}
\subfigure[]{\includegraphics[width=0.5\columnwidth]{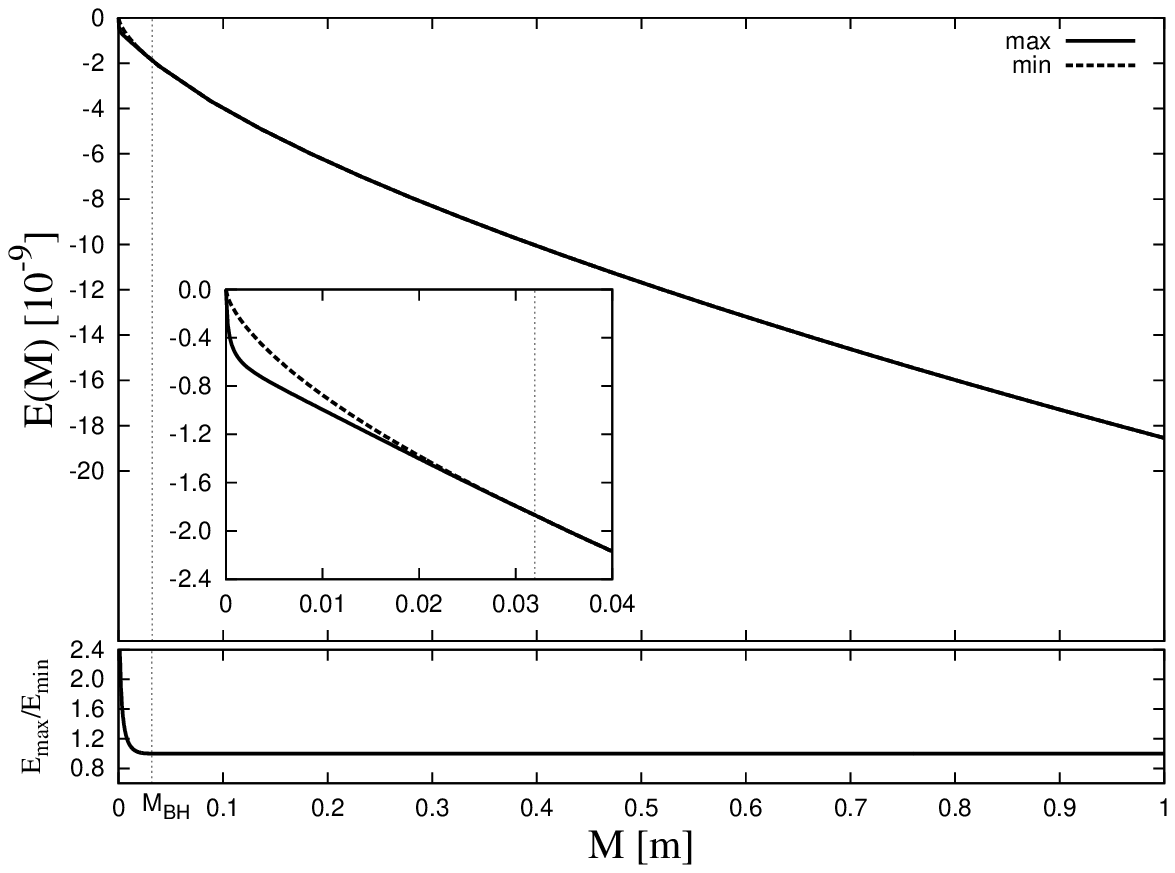}\label{Fig:E_velocity_M87_LCDM}}
\subfigure[]{\includegraphics[width=0.5\columnwidth]{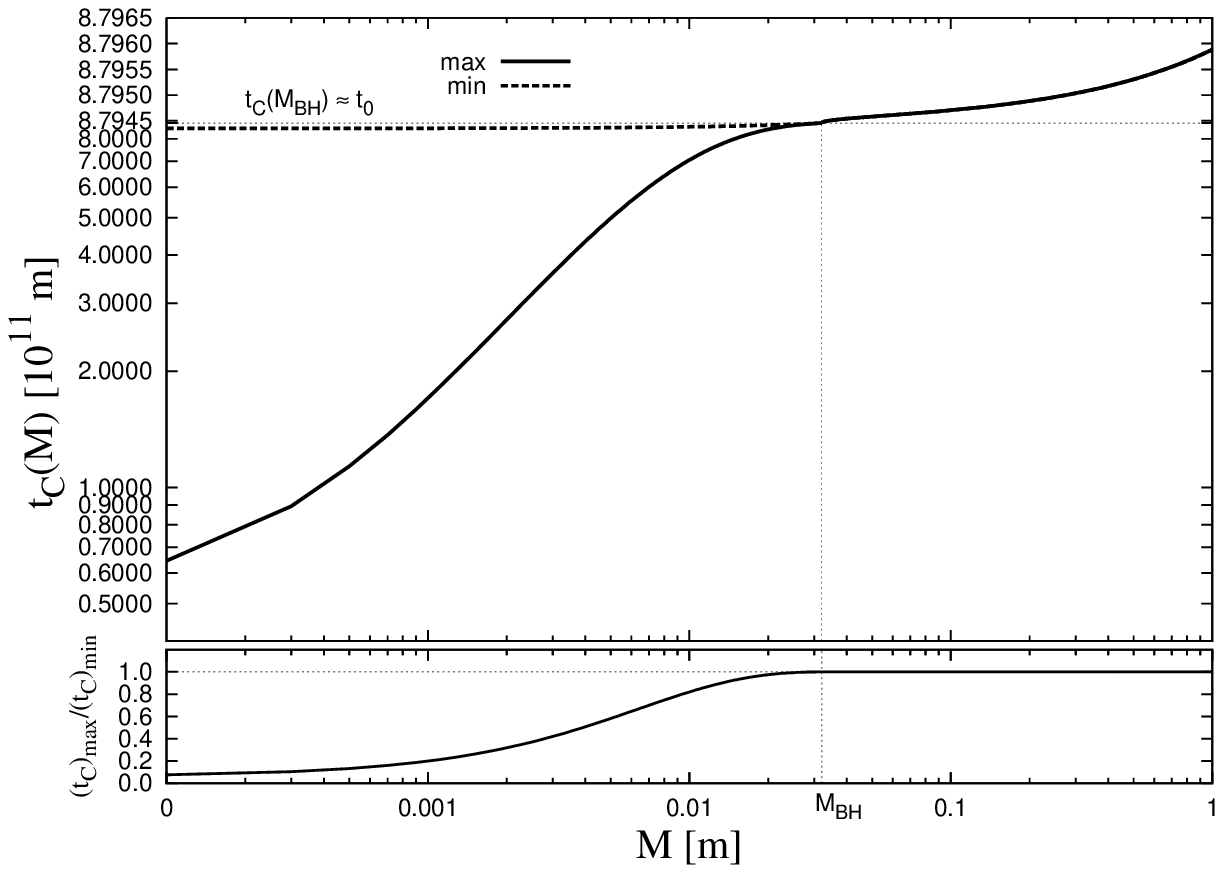}\label{Fig:t_C_velocity_M87_LCDM}}
\caption{\subref{Fig:rho_M_perturbation_velocity_M87_LCDM} The ratio of the
density at the recombination $\rho_{\rm rec}(M)$ corresponding to the
homogeneous velocity profile of the standard $\Lambda$CDM model to the $\Lambda$CDM density in the same model. The nonuniformness of this profile is responsible for
the creation of a galaxy with a central black hole.
\subref{Fig:t_B_velocity_M87_LCDM}--\subref{Fig:E_velocity_M87_LCDM} The main
panel on both plots shows the $t_B(M)$ and $E(M)$ functions for the maximal and
minimal age of the black hole. The maximal age is $12.719$~Gyr, the minimal one
is $0.428$ Gyr. The functions are calculated for an evolution from a homogeneous
velocity profile at the recombination consistent with the $\Lambda$CDM model
with $\Omega_m=0.266$, $\Omega_\Lambda=0.734$. The bang and crunch time
functions inside the black hole are parameterized by the hyperbolic function of
Table~\ref{Tab:BH_interior}. For $M>M_{\rm BH}$ the functions are identical, and
dependent only on the input profiles. The difference is inside the black hole
($M<M_{\rm BH}$) and is exhibited in the inset. The ratio of the two functions
is plotted in the bottom panel. \subref{Fig:t_C_velocity_M87_LCDM} The main
panel shows the $t_C(M)$ function for the LT model of maximal and minimal age of
the black hole. The optimization parameters are the same as described in
\subref{Fig:t_B_velocity_M87_LCDM} and \subref{Fig:E_velocity_M87_LCDM}. Note
the logarithmic scale for the $x$ axis. The $y$ axis is logarithmic up to the
value of $t_C(M_{\rm BH})=8.7945\cdot 10^{11}~\mathrm{m}\approx t_0$, above that
value the scale is linear. This is done in order to make the plot readable, as
there is a significant difference in the black hole's creation time, that is the
value of $t_{\rm AH+}$ at the minimum, which for the maximal model is equal to
$6.458\cdot 10^{10}$~m. The ratio of the two functions is plotted in the bottom
panel. The dotted vertical line in all plots represents the boundary of the
black hole at $M_{\rm BH}=0.032$ m.}\label{Fig:M87_velocity_LCDM_plots-1}
\end{figure}

The current concordance cosmological model is the homogeneous $\Lambda$CDM model with $H_0=71\pm
2.5$~km~s$^{-1}$~Mpc$^{-1}$, $\Omega_m=0.266$ and $\Omega_\Lambda=0.734$, based
on the latest WMAP results \cite{WMAP7:2011a}. In our calculation we used the parameters of this model, described by the equations in Sec.~\ref{Sec:recombination_parameters}, for the initial state. The constructed model of a galaxy with a black hole takes as the input data for
the initial fluctuation at last scattering the profiles of either velocity
$R,_t(M)$, or density $\rho(M)$. Firstly we will present the results of
optimization of black hole's age for the M87 galaxy evolving from the velocity
profile coinciding with that of the $\Lambda$CDM model.
\subsubsection{The velocity profile}

The quantity that is used in our calculations for this evolution type is
$b=R,_t(M)/M^{1/3}$. In an FLRW model it depends only on time, and for $\Omega_m=0.266$, $\Omega_\Lambda=0.734$ its value at recombination is equal to $3.483\cdot 10^{-3}$ m$^{-1/3}$. The homogeneity of the chosen
velocity profile does not prevent the creation of a galaxy as the corresponding
profile of density at last scattering, shown in
Fig.~\ref{Fig:rho_M_perturbation_velocity_M87_LCDM}, is nonuniform, and an
initial fluctuation appears and can gain mass through accretion.

Figures \ref{Fig:t_B_velocity_M87_LCDM}, \ref{Fig:E_velocity_M87_LCDM} and
\ref{Fig:t_C_velocity_M87_LCDM} show the main results of optimalization of the
age of black hole for the evolution from flat velocity profile at recombination.
The bang and crunch time functions were parameterized by the hyperbolic
functions of Table~\ref{Tab:BH_interior}. The black hole of maximal age is
formed at $6.458\cdot 10^{10}$~m, what means that the age of such a black hole
is $12.719$~Gyr. The corresponding values for the minimal age are $8.5197\cdot
10^{11}$~m and $0.428$~Gyr. The last number is roughly equal to the value
obtained in \cite{Krasinski:2004a}. Other parameters of the model are summarized
in Table~\ref{Tab:LCDM_BH_interior_velocity_hyp}.

\begin{table}
\begin{tabular}{c||c|c}
quantity & min & max \\
\hline \hline
BH age [Gyr] & $0.428888$ & $12.719398$\\
$(t_{\rm AH+})_{\rm min}$ [$10^{10}$ m] & $85.197$ & $6.458$\\
$b_1$ [$10^{5}$] & $1.156$ & $1.160$\\
$b_2$ & $374.749$ & $-8.460\cdot 10^{-6}$\\
$b_3$ & $-0.432$ & $64.205$\\
$b_4$ & $0.629$ & $10.819$\\
$c_1$ [$10^{11}$] & $4.121$ & $8.855$\\
$c_2$ & $3.628\cdot 10^{11}$ & $-3.858\cdot 10^{-19}$\\
$c_3$ & $3.163$ & $-153.308$\\
$c_4$ & $0.641$ & $70.526$\\
\end{tabular}
\caption{The parameters of the black hole's bang and crunch time functions for
the models of minimal and maximal age of the black hole. The parameters labeled
as $b_i$ describe the bang time function, $c_i$ describe the crunch time
function. The initial conditions correspond to those of
Fig.~\ref{Fig:M87_velocity_LCDM_plots-1}.}
\label{Tab:LCDM_BH_interior_velocity_hyp}
\end{table}

\begin{figure}
\centering
\includegraphics[width=0.5\columnwidth]{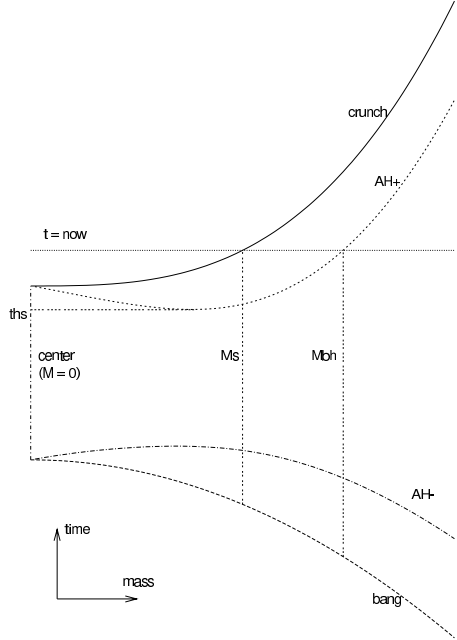}
\caption{Evolution leading to a black hole in the $E<0$ LT model. This a copy of
Fig.~1 from \cite{Krasinski:2004a}, where the bang and crunch time functions
were chosen so as to make the plot readable and illustrative. Intersections of
the line $t={\rm now}$ with the lines representing the Big Crunch and the future
apparent horizon AH+ determine the masses $M_S$ and $M_{\rm BH}$, respectively.
AH-- stands for past apparent horizon and ths for time, when the future apparent
horizon first appears.} \label{Fig:bhspace2}
\end{figure}

As we can see in Fig.~\ref{Fig:t_C_velocity_M87_LCDM} the value of the crunch
time function at $M_{\rm BH}$ is extremely close to $t_0$. However, in the model
the crunch time is equal to $t_0$ not for $M_{\rm BH}$, but for $M_S$, the mass
swallowed by the singularity, which is necessarily smaller than the mass of the
black hole. On the other hand $M_{\rm BH}$ is that value for which the future
apparent horizon crosses the present time. Figure \ref{Fig:bhspace2} shows this
situation in an illustrative way, with the bang and crunch time functions chosen
so as to make the plot readable.

Such behaviour is a consequence of the very small difference between the future apparent
horizon and the crunch time function, which causes $M_S$ and $M_{\rm BH}$ to
almost coincide. The value of $M_S$ cannot be found analytically for a general observational
density profile of the galaxy. In our calculations we have used the method
described in Sec.~\ref{Sec:galaxy_density_profile} that does not require the
knowledge of the value of $M_S$.

\begin{figure}
\subfigure[]{\includegraphics[width=0.5\columnwidth]{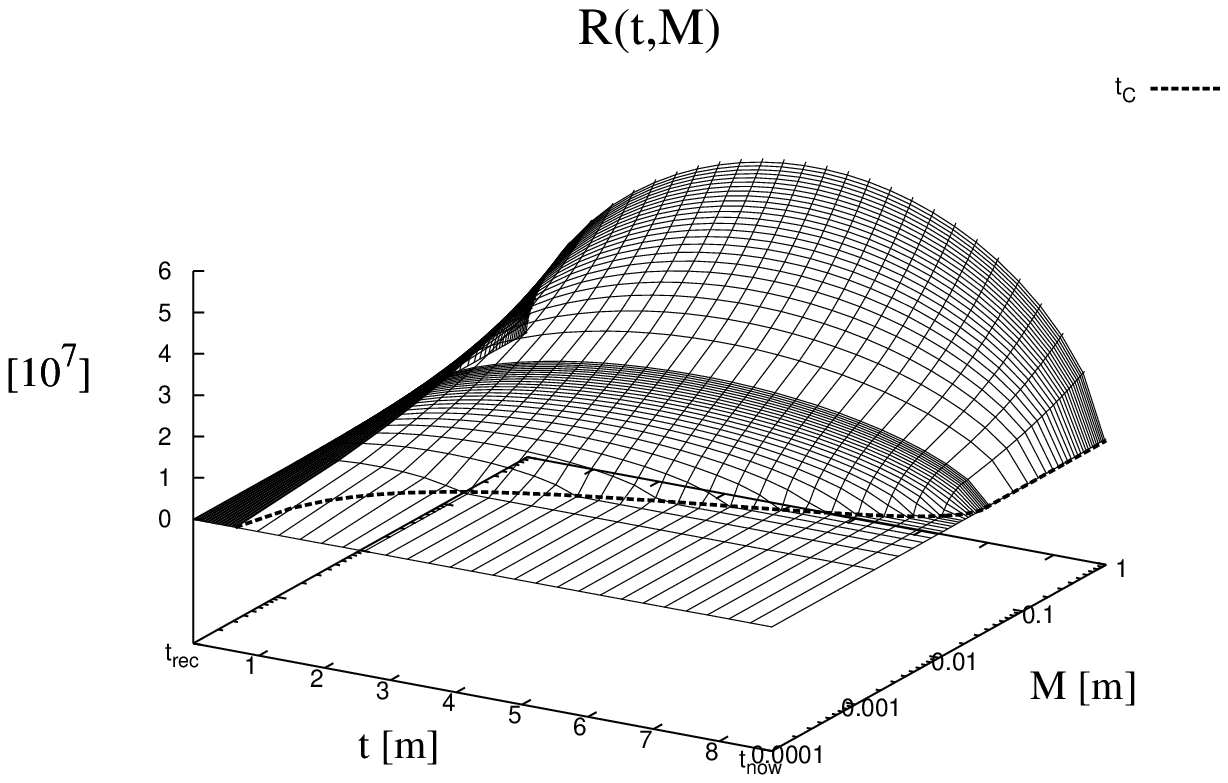}\label{Fig:M87_velocity_LCDM_plots-2-a}}
\subfigure[]{\includegraphics[width=0.5\columnwidth]{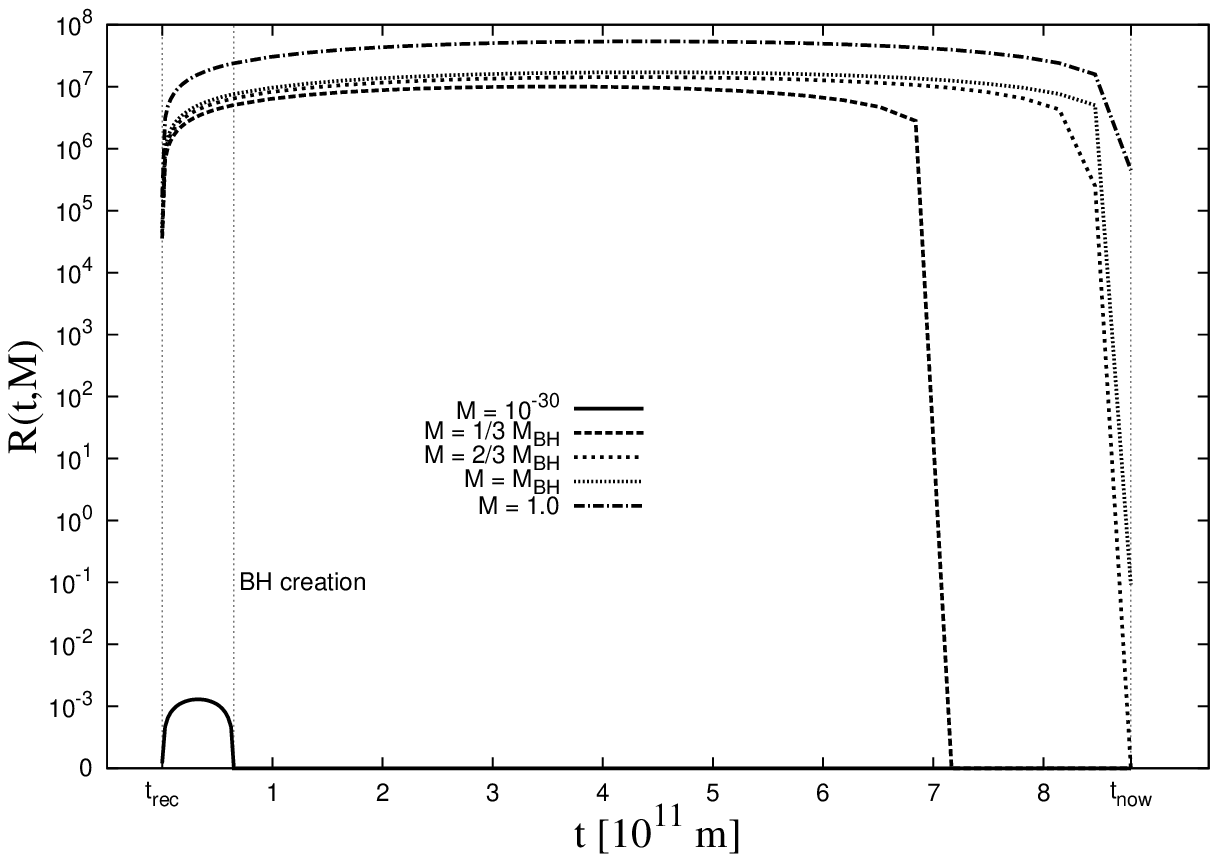}\label{Fig:M87_velocity_LCDM_plots-2-b}}
\subfigure[]{\includegraphics[width=0.5\columnwidth]{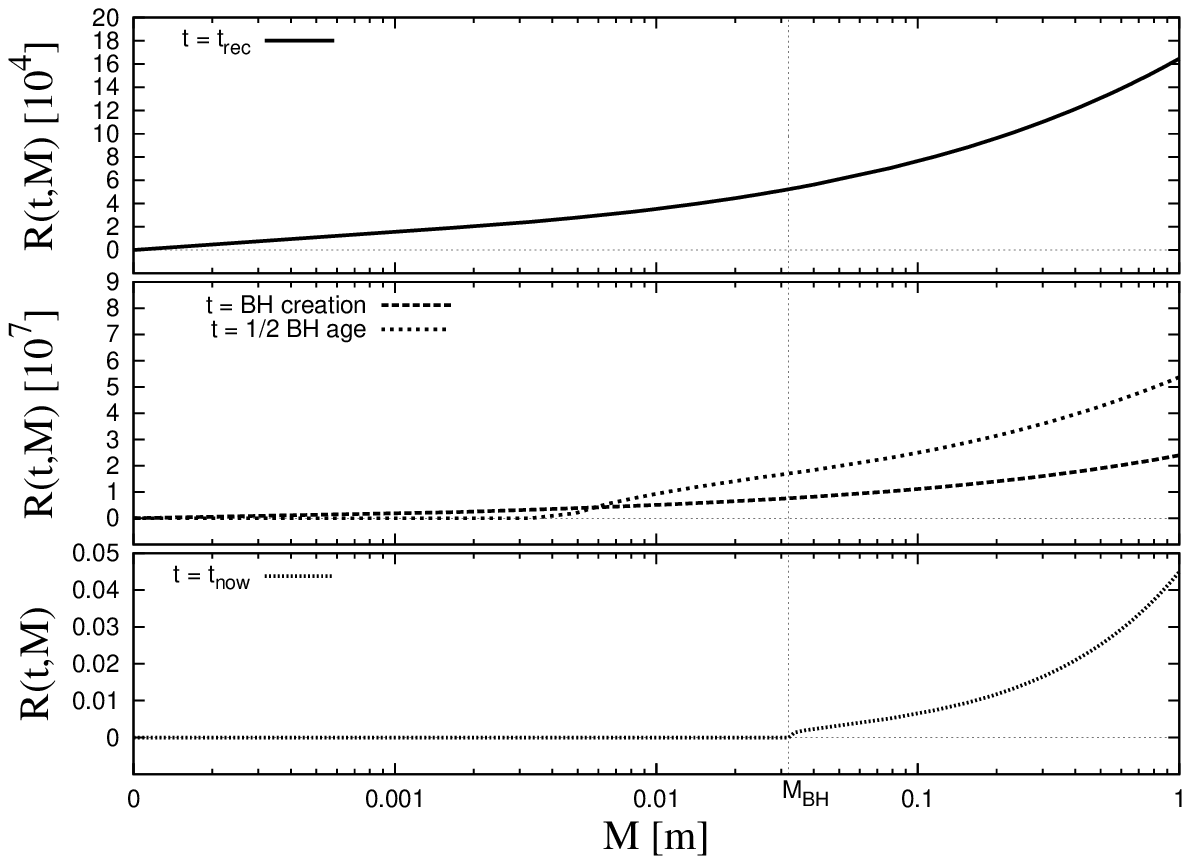}\label{Fig:M87_velocity_LCDM_plots-2-c}}
\caption{\subref{Fig:M87_velocity_LCDM_plots-2-a} The function $R(t,M)$ for the
evolution from flat velocity profile at the recombination. The $M$ axis is in
logarithmic units in order to enhance the region near the center of symmetry.
Note the inscribed crunch time function in the $R=0$ plane - the crunch time
function borders the black hole's interior.
\subref{Fig:M87_velocity_LCDM_plots-2-b} Characteristic $M={\rm const}$ slices
of the areal radius. Only those slices for which $M < M_{\rm BH}$ reach the
black hole. The $M=1$ slice, corresponding to the region furthermost from the
center of symmetry, is recollapsing, but not yet in the final singularity. The
$M=M_{\rm BH}$ slice also does not reach the black hole - its value at $t=t_0$
is $2M_{\rm BH}$. '$t_{\rm rec}$' stands for the time of recombination.
\subref{Fig:M87_velocity_LCDM_plots-2-c} Characteristic $t={\rm const}$ slices
of the areal radius. The upper plot shows the curve for the instant of
recombination, the middle one shows the curves for the time of the creation of
the black hole and for the time when it was half its present age, the bottom one
shows the curve for the present moment. The curves for times after the creation
of the black hole do not start at $M=0$ as the black hole is already
created.}\label{Fig:M87_velocity_LCDM_plots-2}
\end{figure}

The maximal age of the black hole obtained here is in good agreement with the
estimates based on other models. Figure~\ref{Fig:M87_velocity_LCDM_plots-2-a}
shows the time and radial dependence of the areal radius $R(t,M)$. As follows
from the negative value of $E$, the whole region of spacetime under
consideration is recollapsing by now. The flat part of the graph, for which
$t>(t_{\rm AH+})_{\rm min}$, is the region of spacetime where the singularity
already exists.

In Fig.~\ref{Fig:M87_velocity_LCDM_plots-2-b} the constant $M$ slices of
$R(t,M)$ for characteristic $M$ values are shown. The first nonvanishing $M={\rm
const}$ slice is for a small $M$, chosen arbitrarily to be $10^{-30}$. The whole
region under consideration is recollapsing by now, but only the region
$M\leqslant M_{\rm BH}$ is within the black hole. Characteristic $t={\rm const}$ slices are plotted in
Fig.~\ref{Fig:M87_velocity_LCDM_plots-2-c}. The upper plot shows the $t={\rm
const}$ curve for the instant of recombination, the middle one shows the curves
for the time of the creation of the black hole and for the time it was half its
present age, the bottom one shows the curve for the present moment. All the
curves for $t$ values ranging from the recombination time $t_{\rm rec}$ to the
moment of the black hole creation $(t_{\rm AH+})_{\rm min}$ start at the point
$M=0$, $R=0$. All the curves for larger $t$ values start at $M\neq 0$, $R=0$,
because of the formation and growth of the black hole.

\subsubsection{The interior parametrization of the black hole}

\begin{figure}
\subfigure[]{\includegraphics[width=0.5\columnwidth]{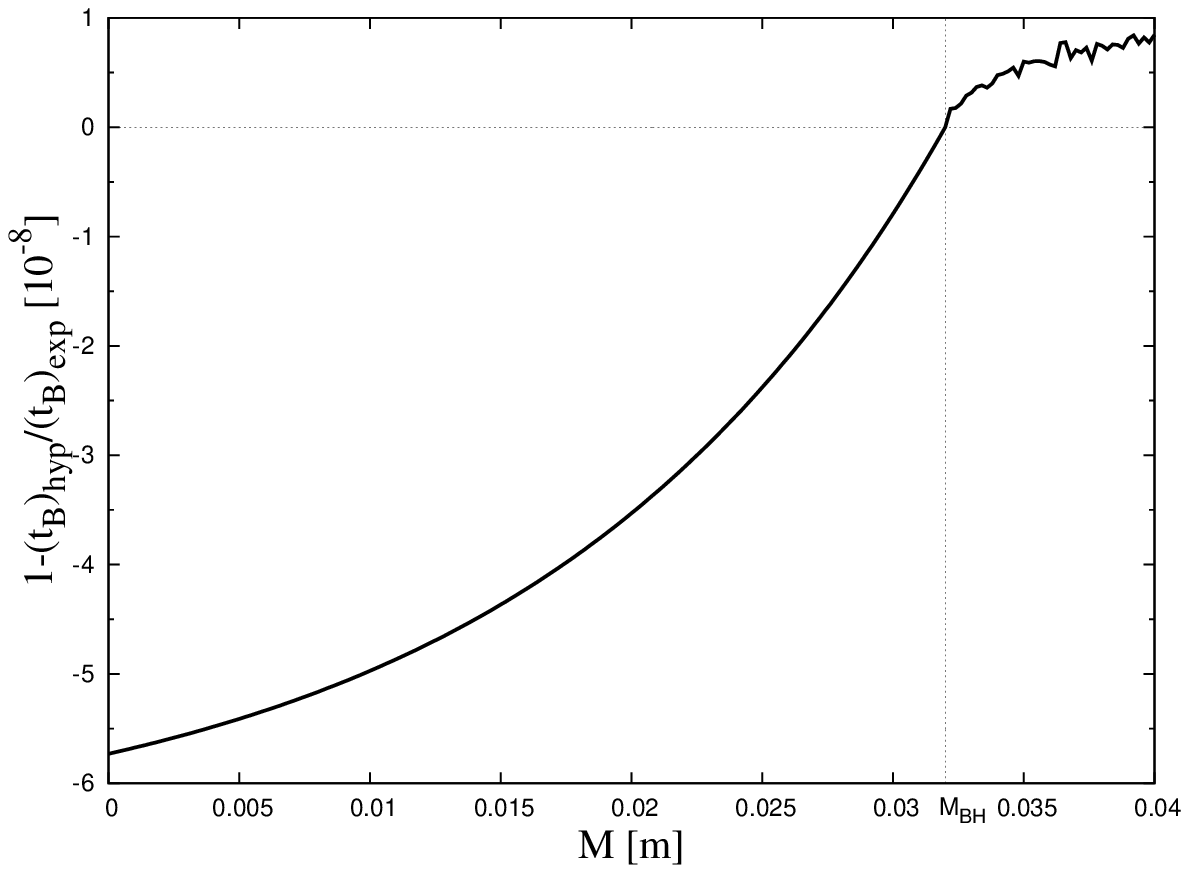}\label{Fig:parametrization_comparison-a}}
\subfigure[]{\includegraphics[width=0.5\columnwidth]{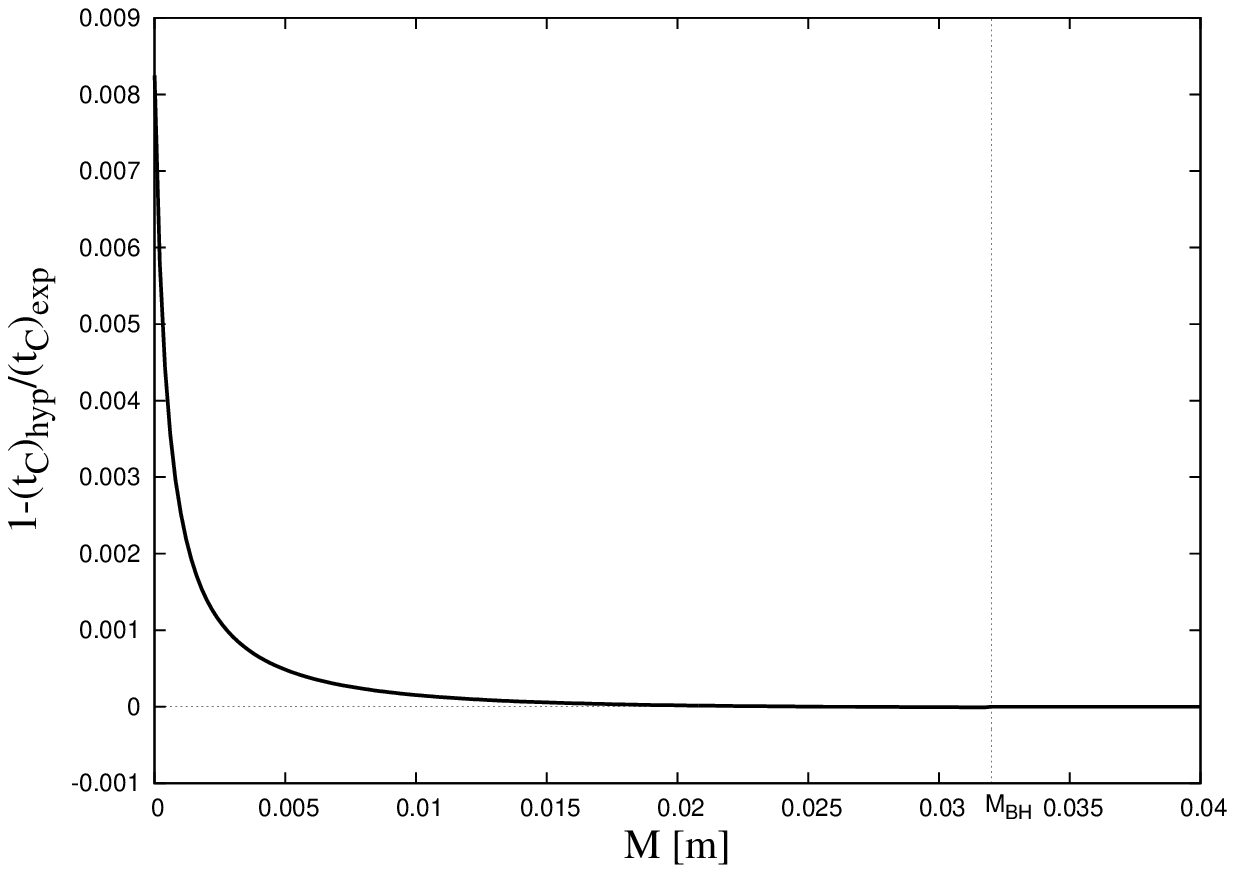}\label{Fig:parametrization_comparison-b}}
\caption{The difference between the hyperbolic and exponential parametrization
of the black hole interior. The plots show the ratio of the bang
\subref{Fig:parametrization_comparison-a} and the crunch
\subref{Fig:parametrization_comparison-b} time functions subtracted from unity
in order to make the values readable. The difference between the functions can
be neglected. The shape of the bang time curve for $M>M_{\rm BH}$ is due to the
numerical errors, as this $M$ range corresponds to the exterior of the black
hole where the type of parametrization of the interior does not affect the
values of the functions.} \label{Fig:parametrization_comparison}
\end{figure}

All the results presented so far were found for the bang and crunch time
functions of the black hole interior given by the hyperbolic function of
Table~\ref{Tab:BH_interior}. However, the calculations can be performed for an
arbitrary type of the interior  of the black hole. Figure
\ref{Fig:parametrization_comparison} presents the comparison between the
hyperbolic and the exponential parametrization. The bang and crunch time
functions are very similar, their ratio differs from unity by a number of the
order of $10^{-8}$ for the bang and $10^{-3}$ for the crunch time function. The
maximal age of the black hole is almost the same for the two types of
parametrization. Therefore in the following we have chosen only the hyperbolic
parametrization of the interior of the black hole functions.
\subsubsection{The density profile}

An initial fluctuation at the recombination can also be defined by a density
profile, that is the function $\rho_{\rm rec}(M)$. The quantity that is used
here in calculations is $a=R/M^{1/3}$, where the areal radius at the
recombination is found using the modified Eq.~\eqref{Eq:R_now_galaxy}, without
the first term in square brackets and setting the lower limit of the integral to
zero. As in the evolution from a given velocity profile at last scattering, we
set the density profile to the one corresponding to the $\Lambda$CDM model with
$\Omega_m=0.266$ and $\Omega_\Lambda=0.734$. Then the density profile is flat
and the value of $a$ is equal to $1.649\cdot 10^{5}$~m$^{-1/3}$.

\begin{figure}
\subfigure[]{\includegraphics[width=0.5\columnwidth]{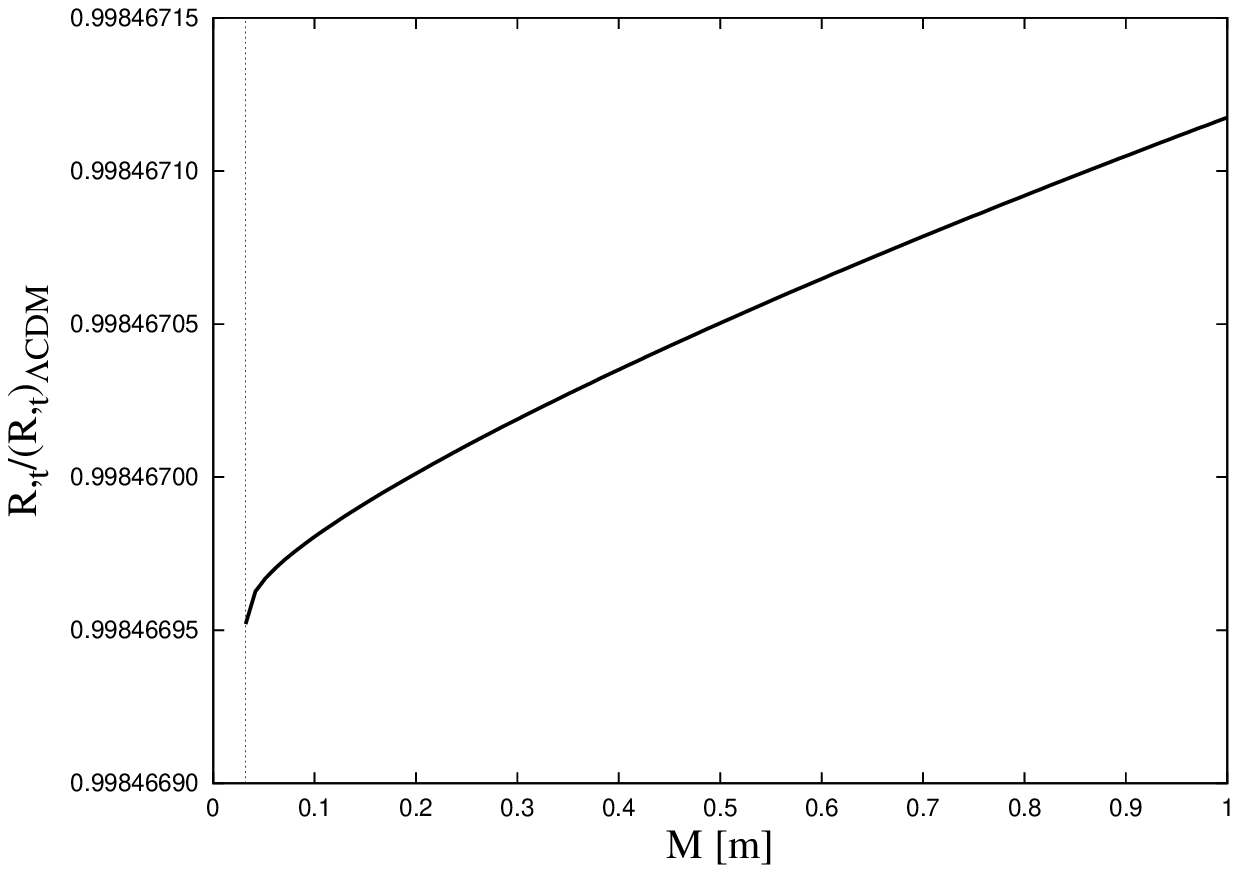}\label{Fig:M87_density_LCDM_plots-1}}
\subfigure[]{\includegraphics[width=0.5\columnwidth]{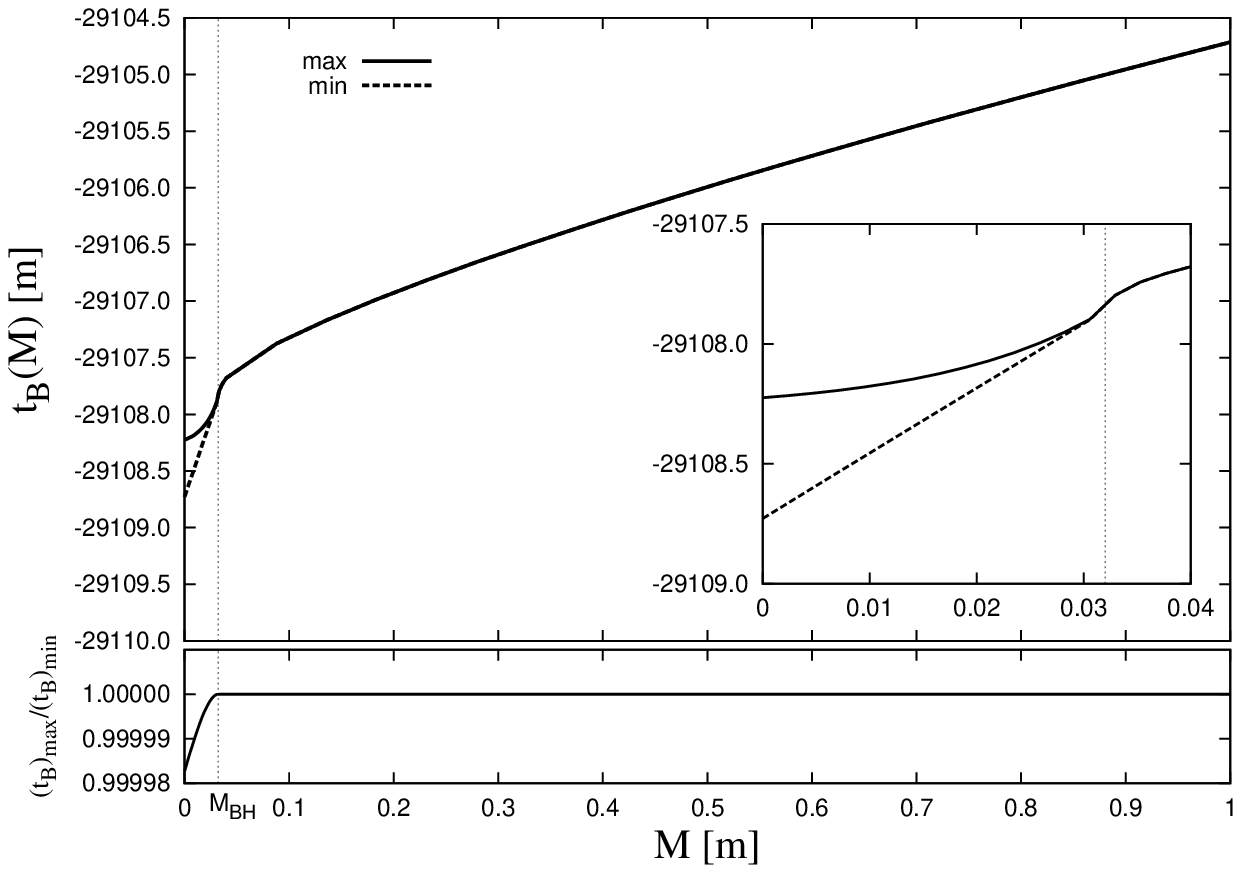}\label{Fig:t_B_density_M87_LCDM}}
\subfigure[]{\includegraphics[width=0.5\columnwidth]{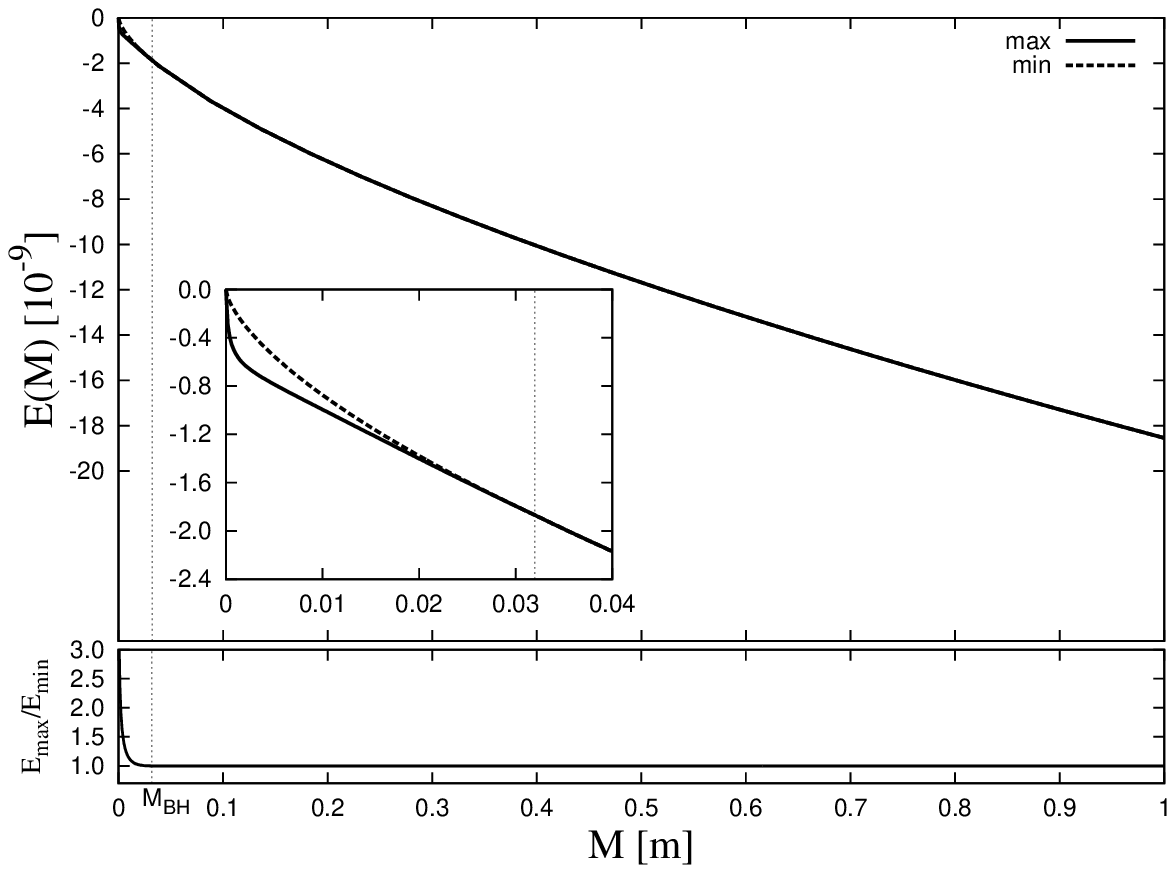}\label{Fig:E_density_M87_LCDM}}
\subfigure[]{\includegraphics[width=0.5\columnwidth]{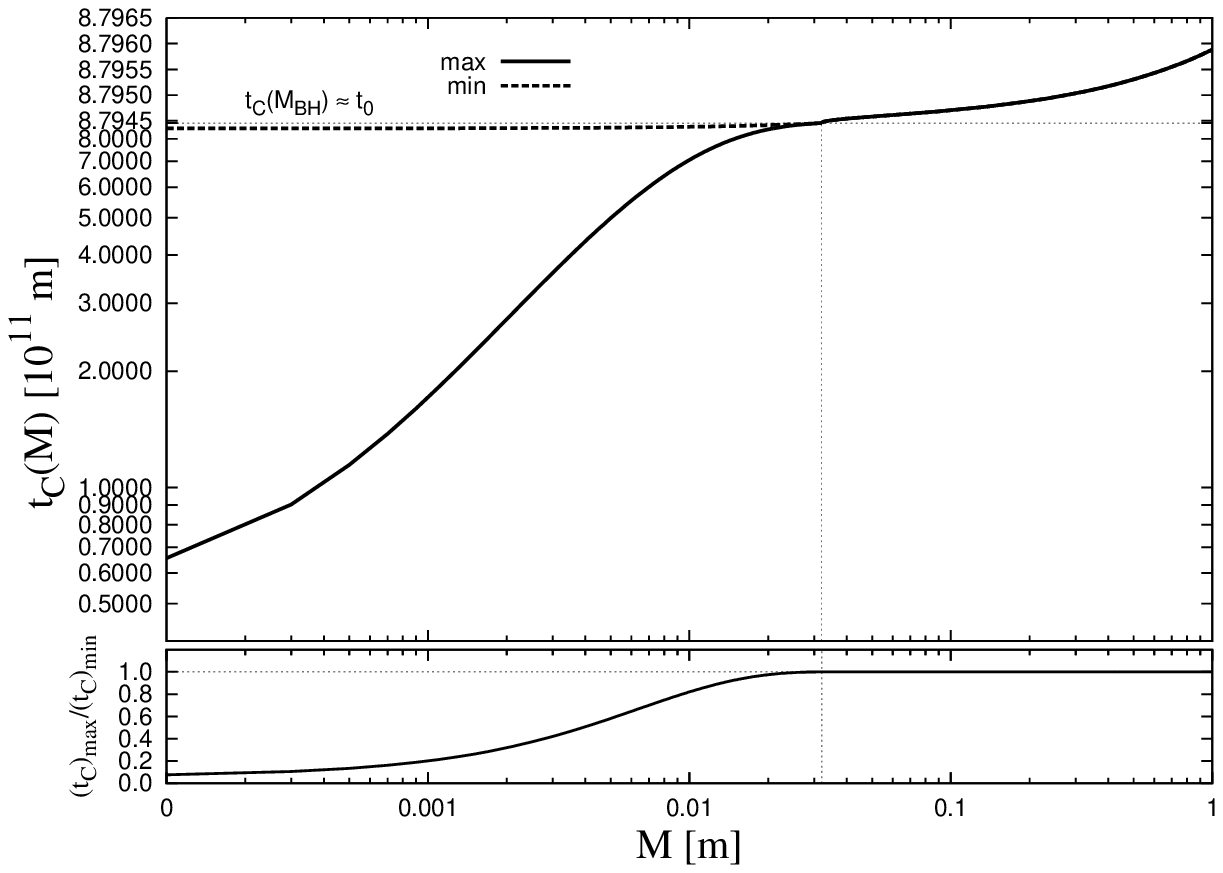}\label{Fig:t_C_density_M87_LCDM}}
\caption{\subref{Fig:M87_density_LCDM_plots-1} The ratio of the velocity at last
scattering corresponding to a flat density profile, to the $\Lambda$CDM velocity. A monotonic increase of this function enables the
creation of a galaxy with a black hole. \subref{Fig:t_B_density_M87_LCDM} The
bang time function for the evolution from the flat density profile for the
maximal and minimal age of the black hole models. The main panel shows the
functions for the whole $M$ range, the inset shows the functions near the center
of symmetry at $M=0$ and the bottom one shows the ratio of the maximal model to
the minimal. Note the difference from the bang time function for the evolution from flat velocity profile in
Fig.~\ref{Fig:t_B_velocity_M87_LCDM}. This function being increasing indicates
that there is a shell crossing in the model, but it appears outside the range of
the application of the model. \subref{Fig:E_density_M87_LCDM} The corresponding
$E(M)$ function in the same range as for the velocity evolution in
Fig.~\ref{Fig:E_velocity_M87_LCDM}. The difference between this and the flat
velocity evolution is very small. \subref{Fig:t_C_density_M87_LCDM} In the main
panel: the crunch time function for the maximal an minimal black hole models.
The age of the black hole for the maximal model is $12.719$~Gyr, and $0.429$~Gyr
for the minimal one, the same values (within numerical errors) as for the
evolution from a flat velocity profile. Note the logarithmic $x$
axis.}\label{Fig:M87_density_LCDM_plots-2}
\end{figure}

The corresponding profile of velocity $R,_t(M)$ is presented in
Fig.~\ref{Fig:M87_density_LCDM_plots-1}. This profile exhibits nonuniformness
enabling the creation of a black hole. However, the nature of the fluctuation is
different than in the case of a flat velocity profile and its corresponding
density profile in Fig.~\ref{Fig:rho_M_perturbation_velocity_M87_LCDM}. In the
model the density at the center increases by accretion of background matter,
which is achieved by a lower expansion rate near the center and a higher rate
further away. The difference in expansion rates is the underlying reason of
accretion and the creation of black hole. Therefore, any velocity profile at
last scattering that could lead to the creation of a galaxy with a black hole
must necessarily be monotonically increasing as a function of $M$. Such a
behaviour can be seen in Fig.~\ref{Fig:M87_density_LCDM_plots-1}.

\begin{figure}
\subfigure[]{\includegraphics[width=0.5\columnwidth]{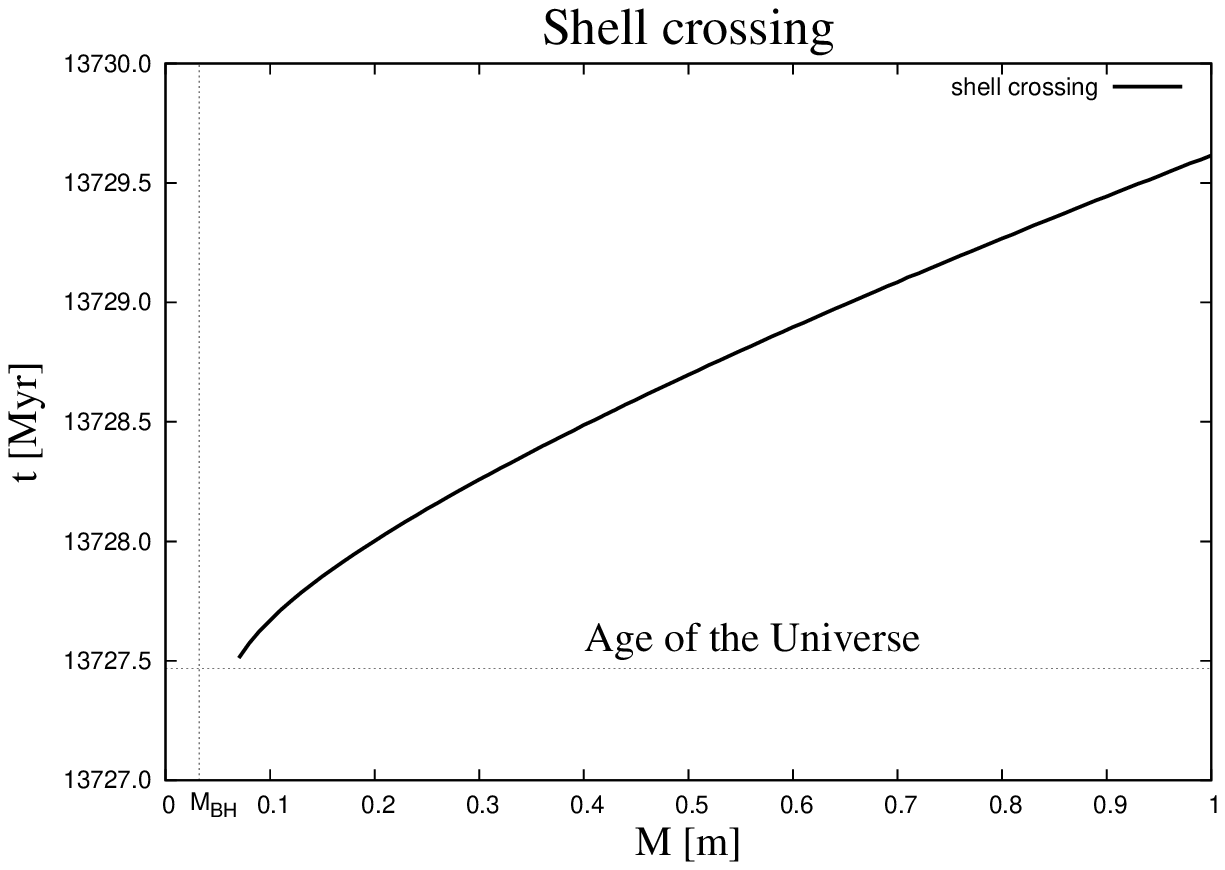}\label{Fig:shell_crossing_LCDM}}
\subfigure[]{\includegraphics[width=0.5\columnwidth]{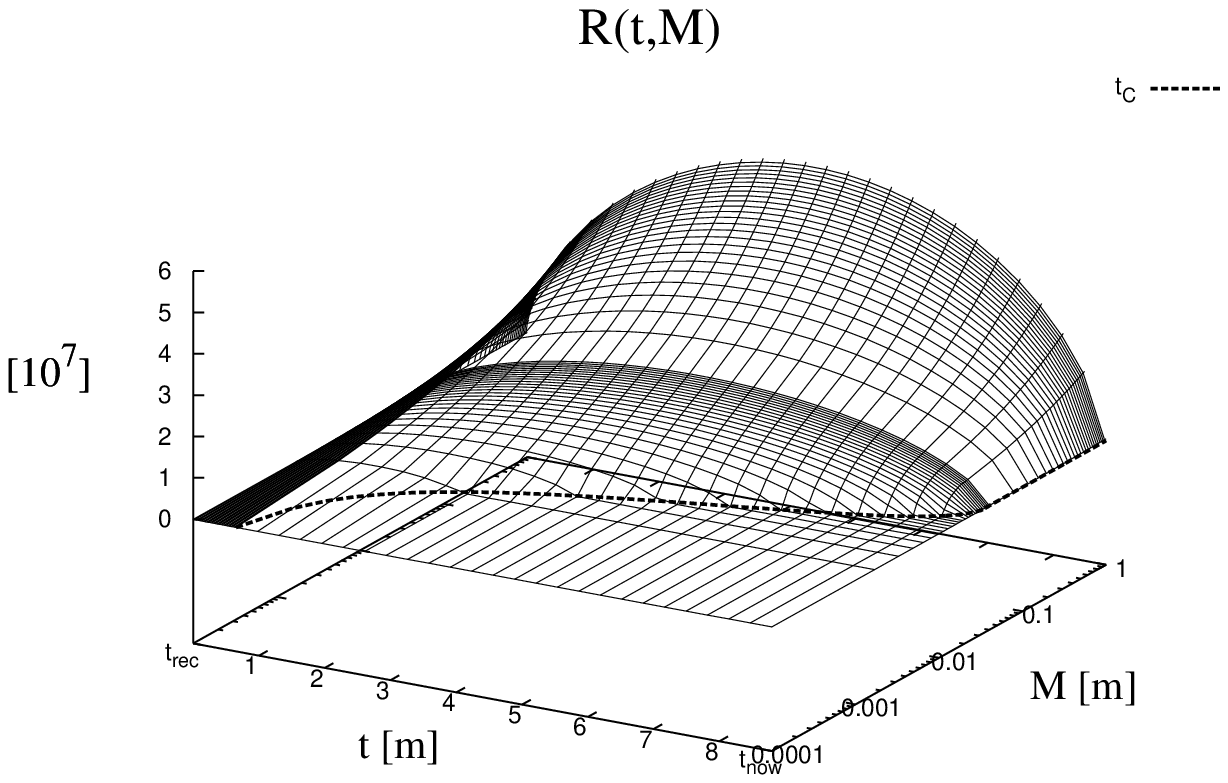}\label{Fig:R_3D_density_M87_LCDM}}
\caption{\subref{Fig:shell_crossing_LCDM} The coordinates of the points where
the shell crossing singularity occurs. The shell crossing appears after the
present moment, i.e. not in the range of application of the model. \subref{Fig:R_3D_density_M87_LCDM}
Evolution of the areal radius $R(t,M)$ for the evolution from a flat density profile at the
recombination. There are no shell crossings as there are no points (besides the
black hole interior) where $R,_M=0$.}\label{Fig:M87_density_LCDM_plots-3}
\end{figure}

The results of our calculations for the evolution form a flat density profile
coinciding with the $\Lambda$CDM profile are presented in
Figs.~\ref{Fig:t_B_density_M87_LCDM}--\ref{Fig:t_C_density_M87_LCDM}. The $E(M)$
functions are similar in both cases and the whole $M$ range is recollapsing
which is necessary for the creation of a black hole. The main difference
compared to the previous evolution type is in the bang time function $t_B(M)$
which is increasing. This means that the conditions of avoidance of shell
crossings are not fulfilled and a shell crossing is imminent. However, due to
the values of $t_B(M)$ it will occur after the present time and not in the range
of application of the model. Figure~\ref{Fig:shell_crossing_LCDM} shows the
points where the shell crossings appear. We can see that for the points near the
black hole it appears after the present moment. The non--occurrence of the shell
crossing in the area covered by the model can also be seen in
Fig.~\ref{Fig:R_3D_density_M87_LCDM} where there are no points (besides the
black hole interior) where $R,_M=0$. The small difference in the LT model
functions means that the ages of black holes in both evolution scenarios are the
same, within the numerical errors. Other parameters of the bang and crunch time
functions are shown in Table~\ref{Tab:LCDM_BH_interior_density_hyp_comparison}.

\begin{table}
\begin{tabular}{c||c|c}
quantity & min & max\\
\hline \hline
BH age [Gyr] & $0.429112$ & $12.719378$ \\
$(t_{\rm AH+})_{\rm min}$ [$10^{10}$ m] & $8.5195$ & $6.4583$\\
$b_1$ [$10^{4}$] & $-2.890$ & $-2.911$\\
$b_2$ & $-94.851$ & $2.141\cdot 10^{-6}$\\
$b_3$ & $-0.432$ & $64.205$\\
$b_4$ & $0.629$ & $10.819$\\
$c_1$ [$10^{11}$] & $4.119$ & $8.855$\\
$c_2$ & $3.630\cdot 10^{11}$ & $-3.858\cdot 10^{-19}$\\
$c_3$ & $3.163$ & $-153.308$\\
$c_4$ & $0.641$ & $70.526$\\
\end{tabular}
\caption{The parameters of the bang and crunch time functions of the interior of
the black hole for the LT models of minimal and maximal age of the black hole
for evolution from a flat density profile coinciding with the standard $\Lambda$CDM model. As before, the parameters labeled as $b_i$ describe
the bang time function, $c_i$ describe the crunch time function.}
\label{Tab:LCDM_BH_interior_density_hyp_comparison}
\end{table}

\subsection{The parameters at the recombination}

\afterpage{
\begin{figure}
\includegraphics[width=0.5\columnwidth]{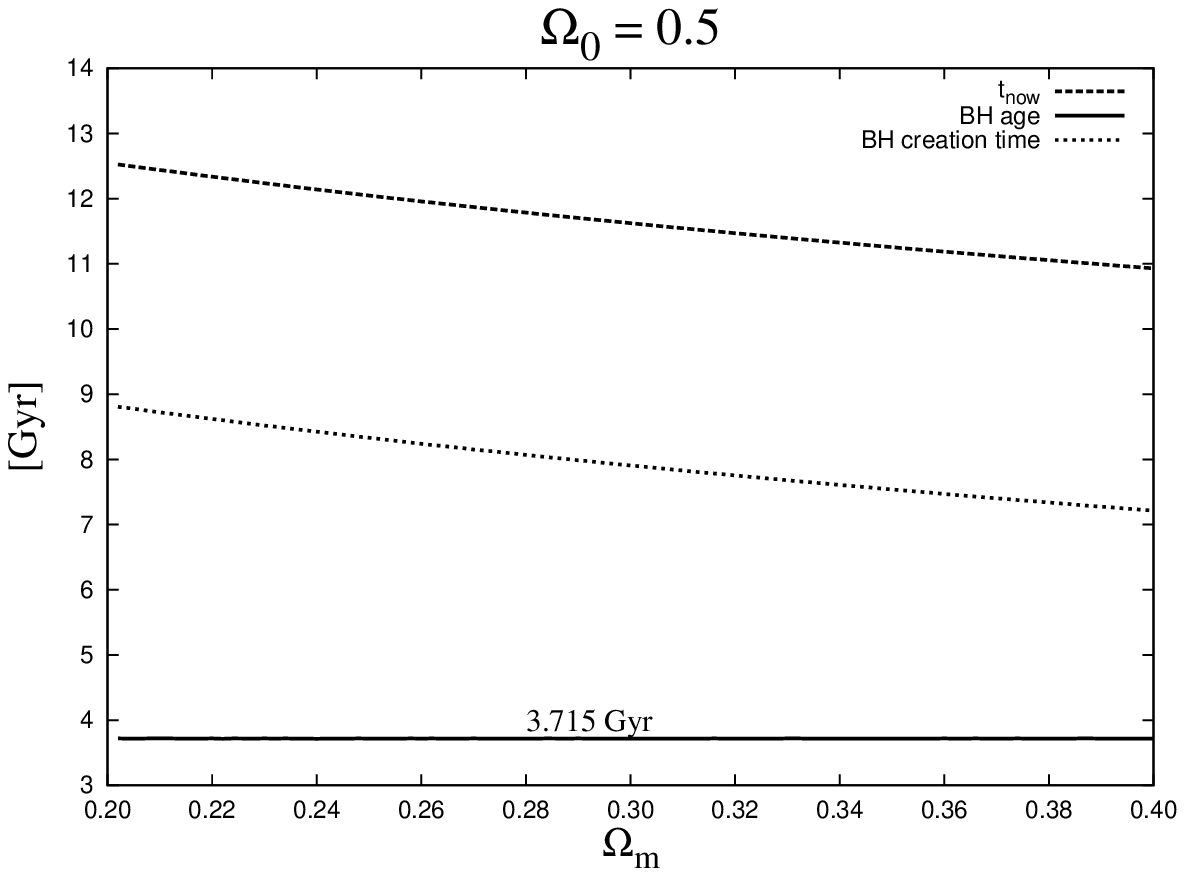}
\includegraphics[width=0.5\columnwidth]{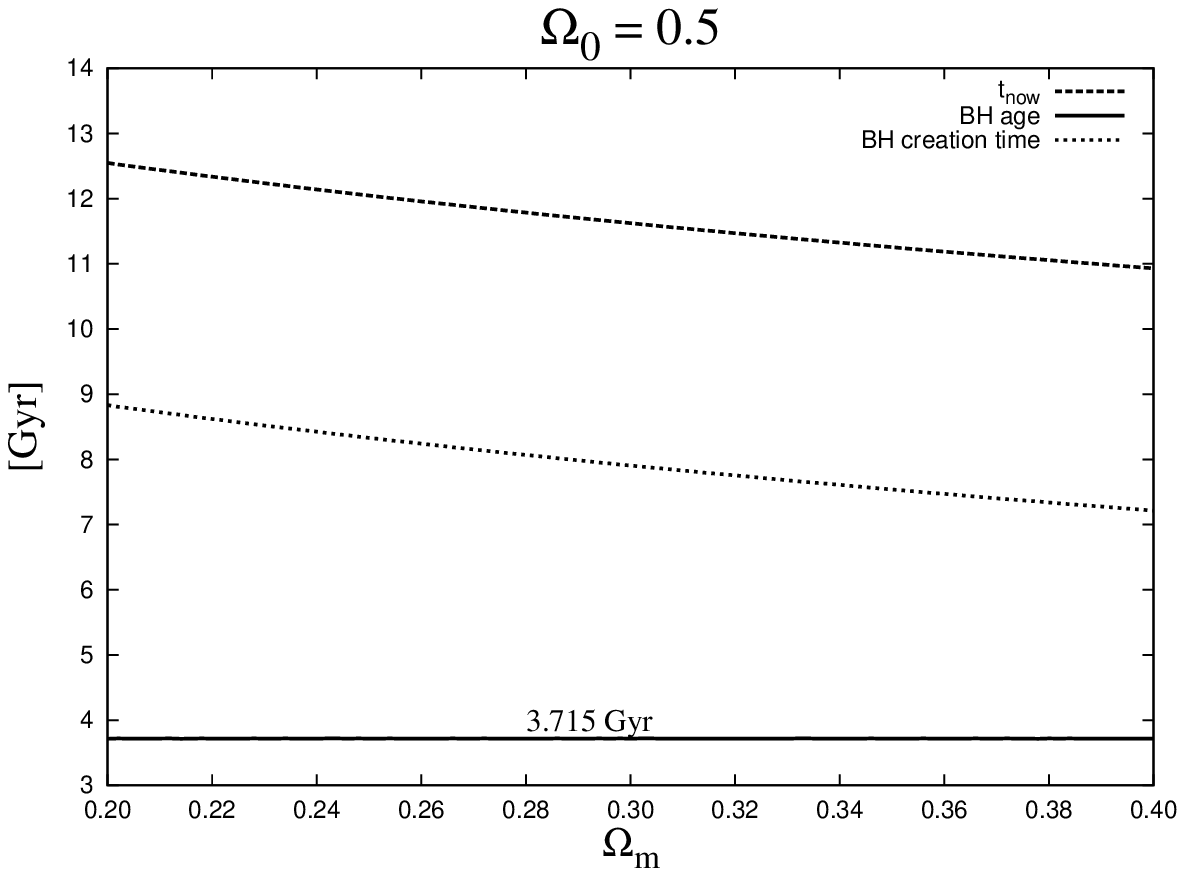}
\\
\includegraphics[width=0.5\columnwidth]{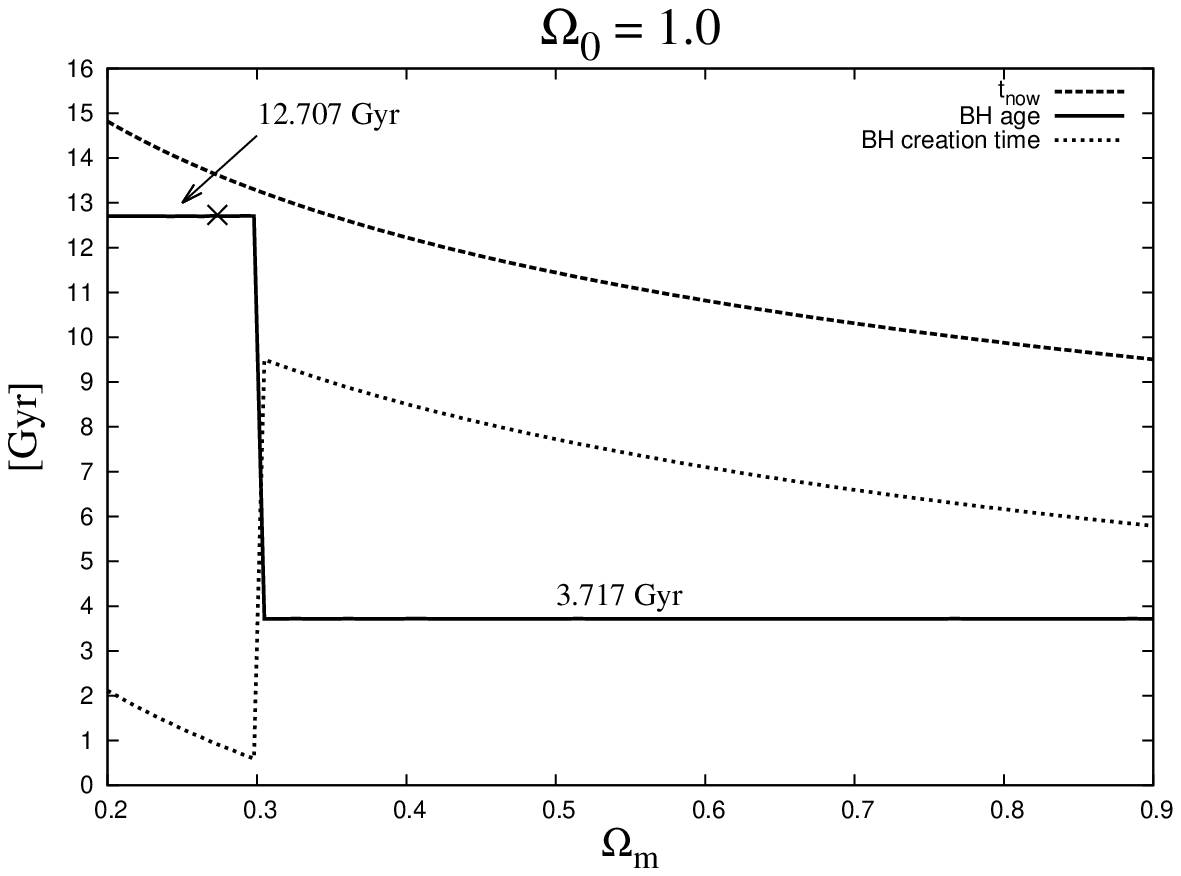}
\includegraphics[width=0.5\columnwidth]{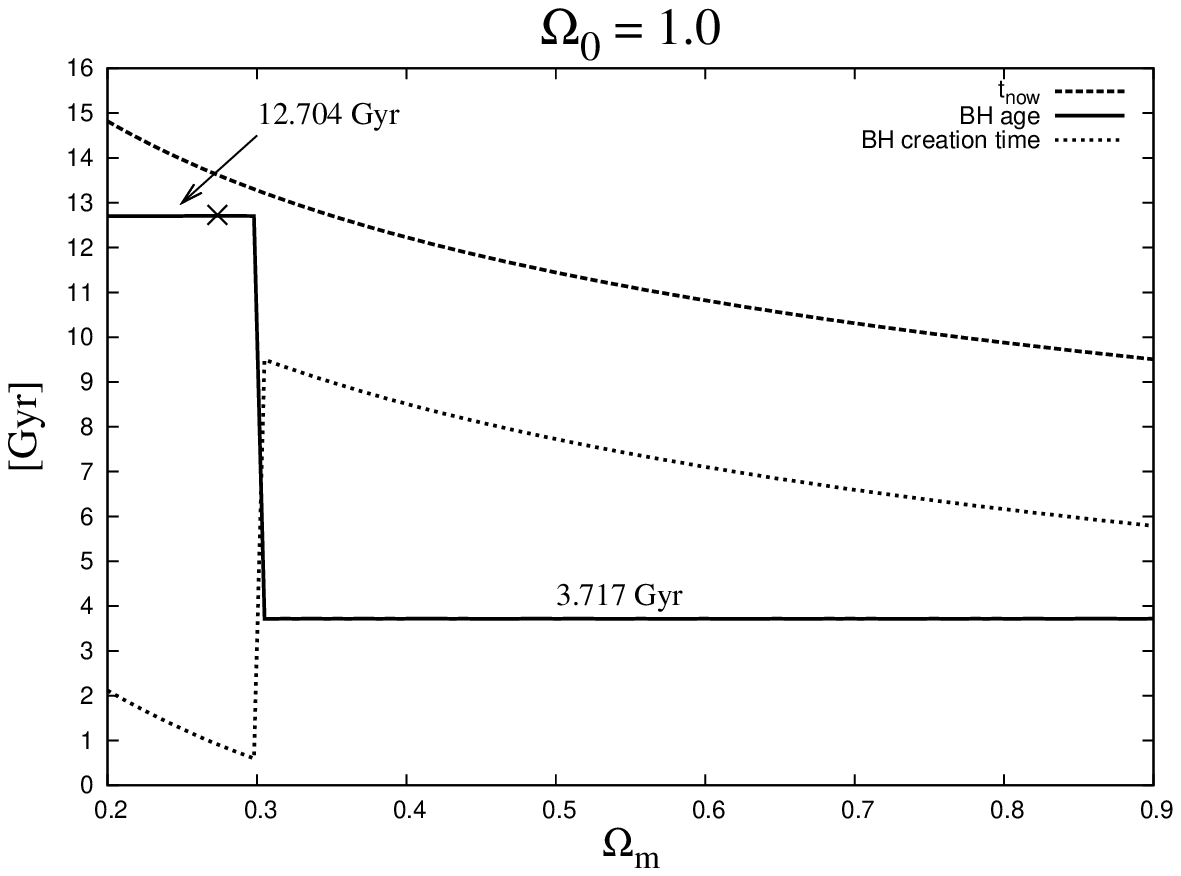}
\\
\includegraphics[width=0.5\columnwidth]{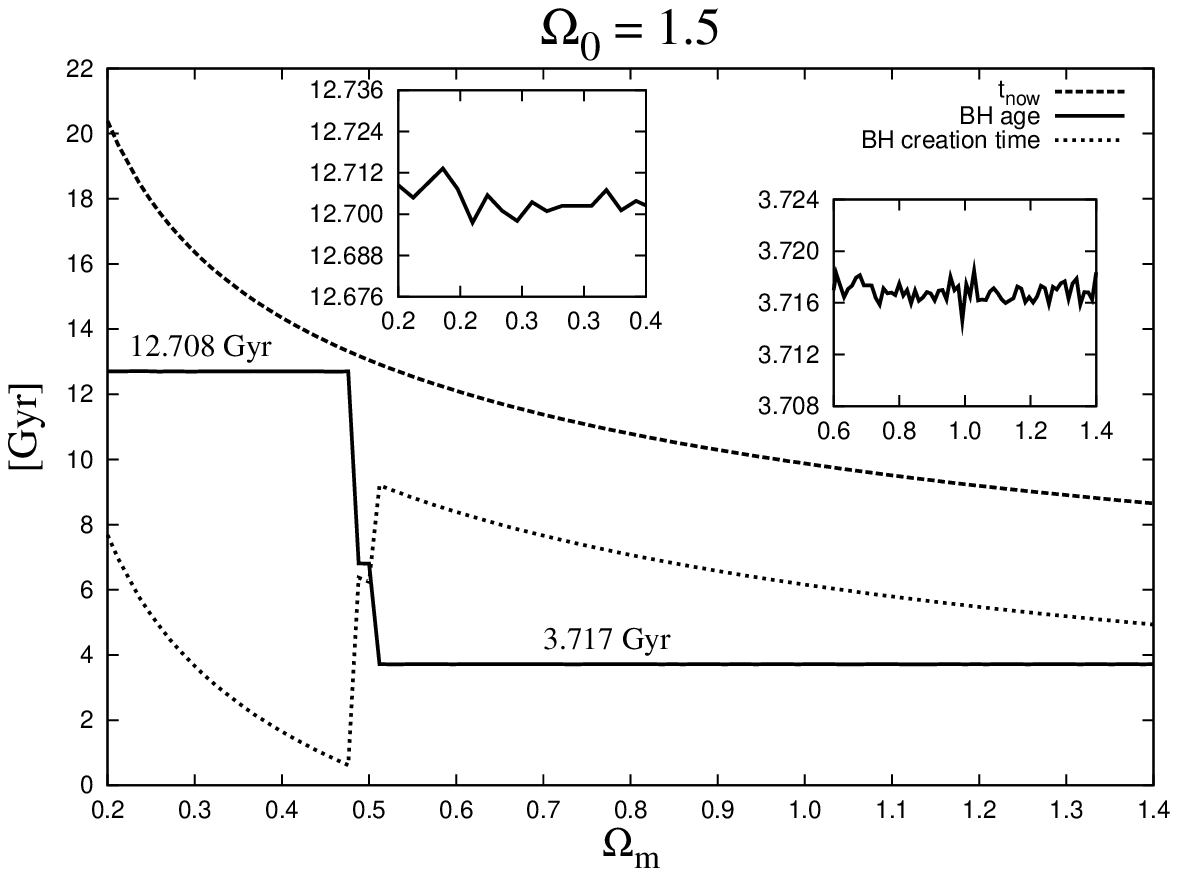}
\includegraphics[width=0.5\columnwidth]{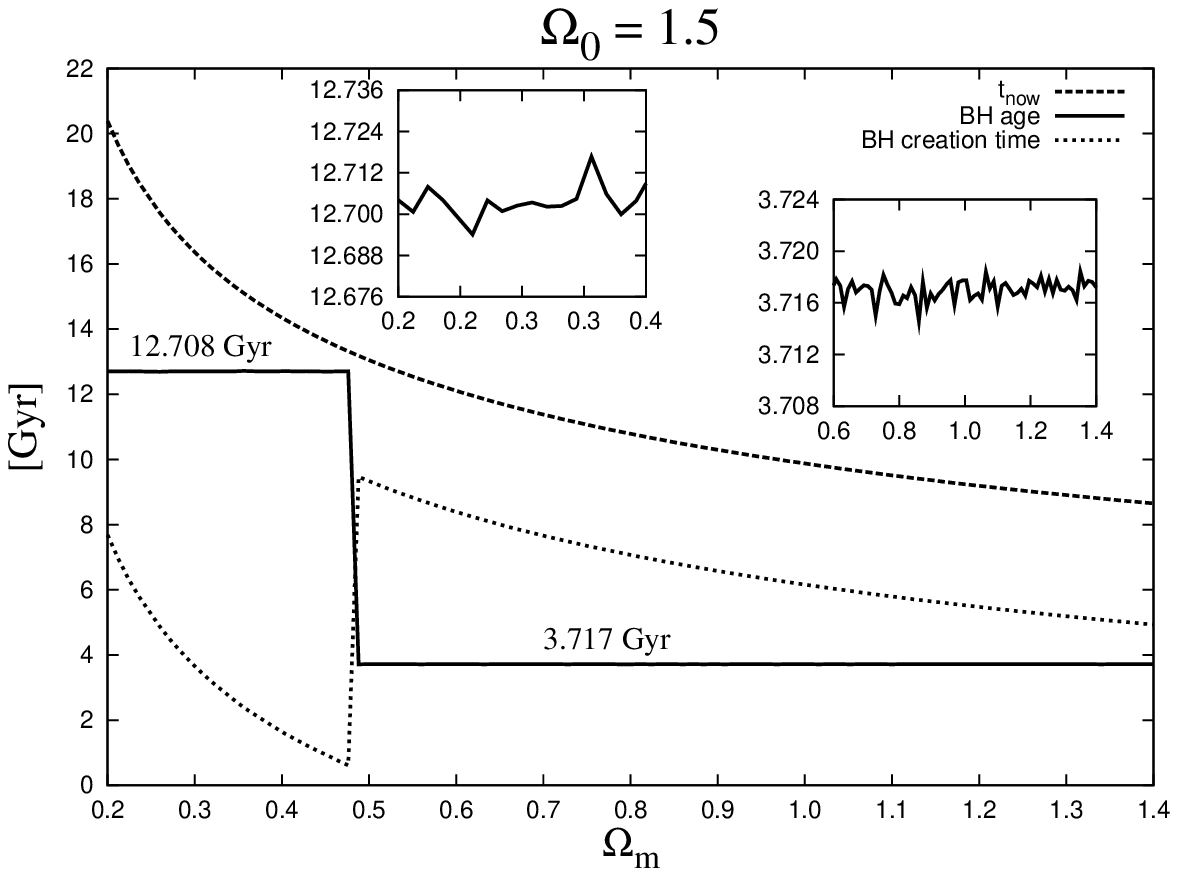}
\caption{The maximal age of the black hole for three different values of
$\Omega_0$ for a flat initial velocity profile (left) and a flat initial density
profile (right) as a function of $\Omega_m$. The interior of the black hole is
parameterized using hyperbolic functions. The two constant values of the maximal
age of the black hole are clearly visible. In the $\Omega_0=0.5$ case, for the
whole range of $\Omega_m$ values, only $3.717$~Gyrs old black holes are formed.
For the other two values of $\Omega_0$ also $12.708$~Gyrs old black holes are
created. In the $\Omega_0=1.0$ plots the cross indicates the $\Lambda$CDM model.
The insets in the $\Omega_0=1.5$ plots show the small changes of the age values,
caused by numerical approximations and the usage of random numbers for initial
conditions.} \label{Fig:const_omega_zero_analysis-BH_age}
\end{figure}
}

\afterpage{
\begin{figure}
\includegraphics[width=0.5\columnwidth]{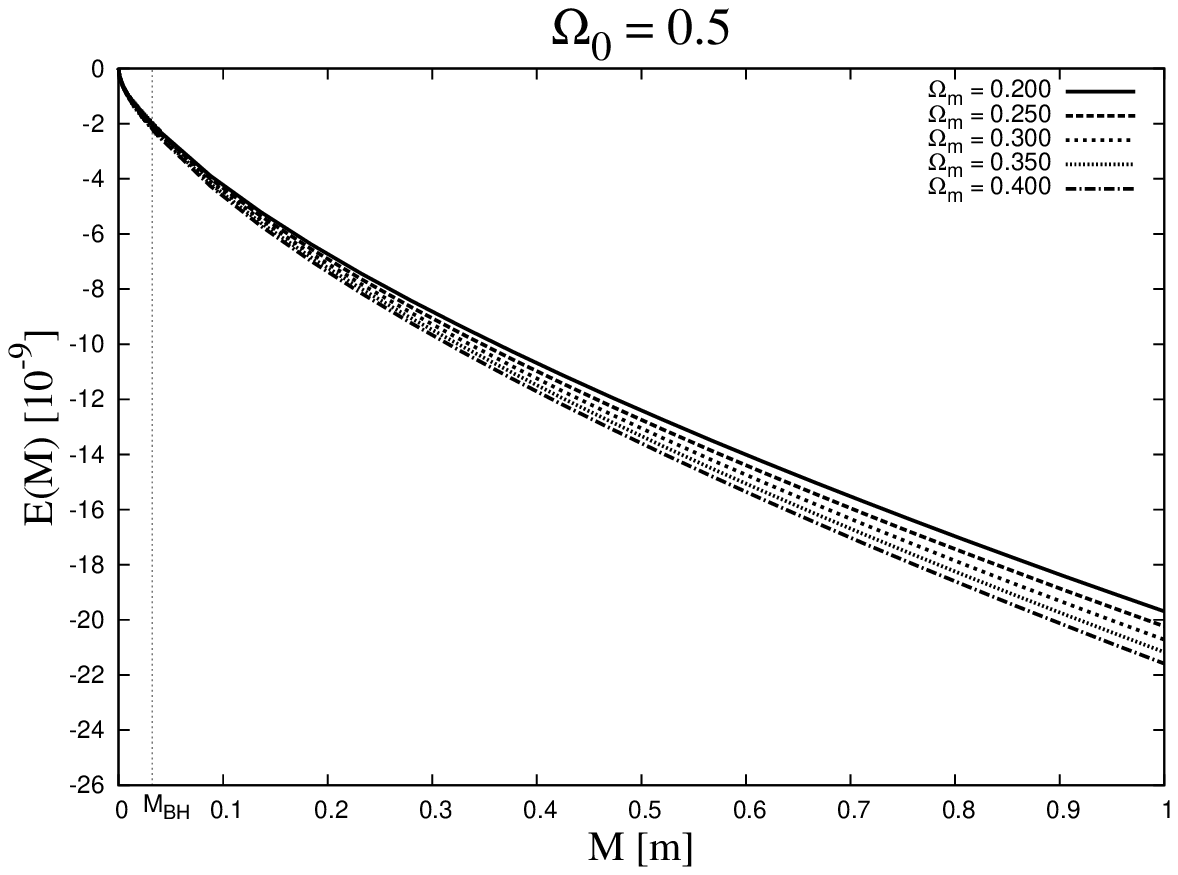}
\includegraphics[width=0.5\columnwidth]{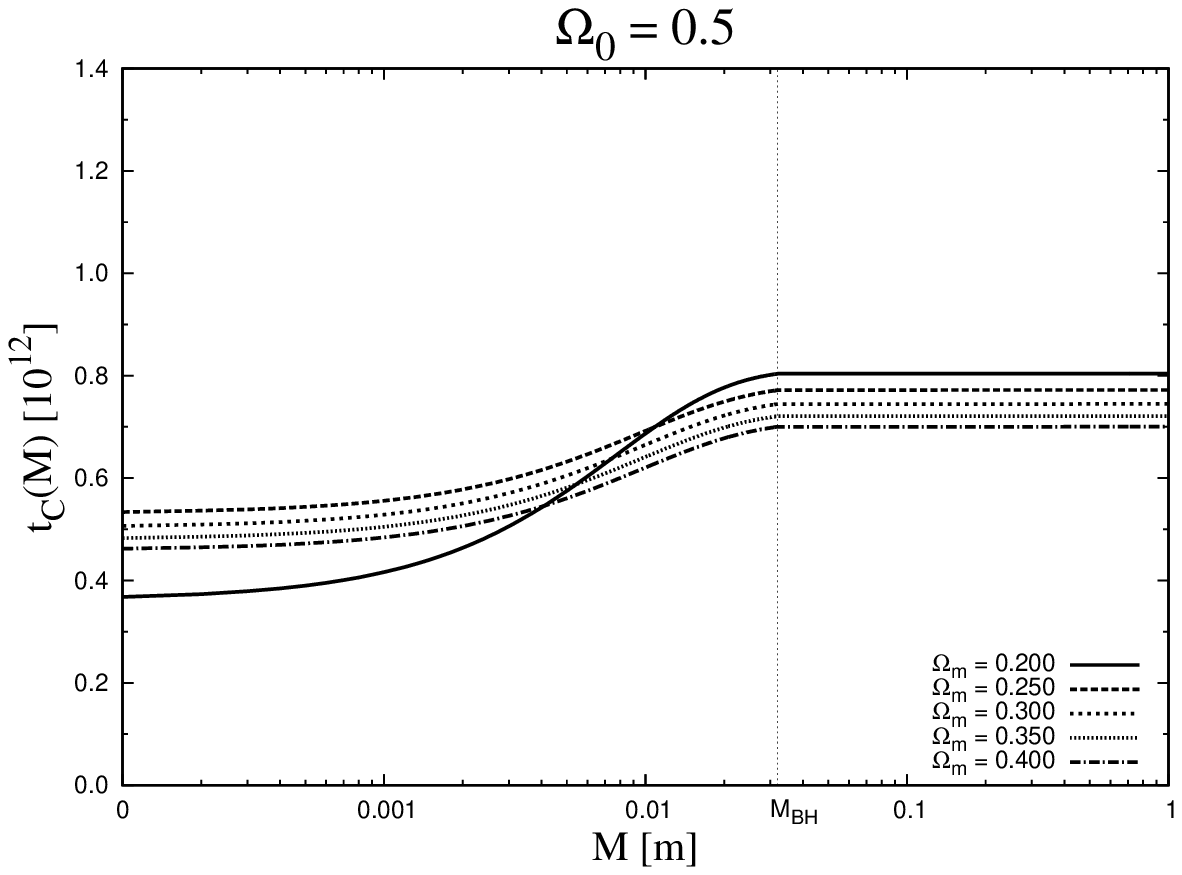}
\\
\includegraphics[width=0.5\columnwidth]{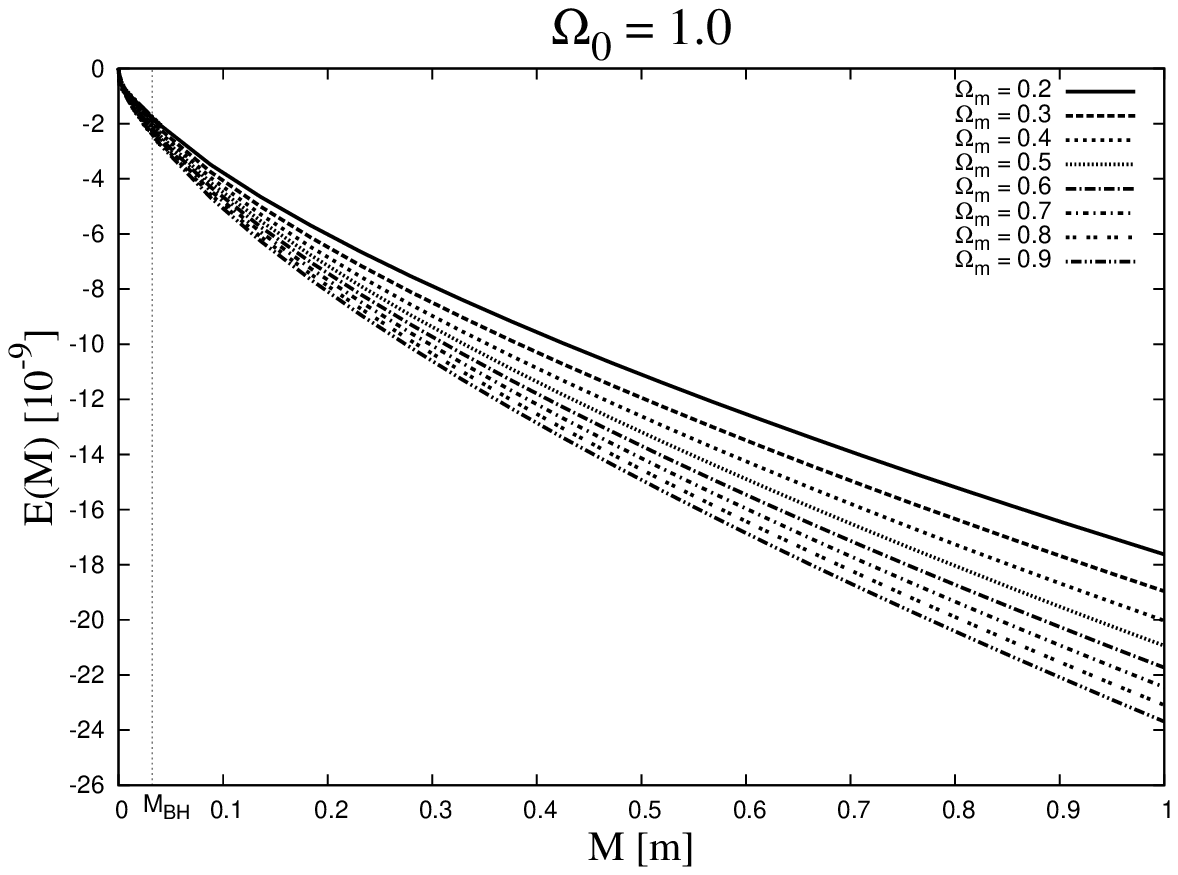}
\includegraphics[width=0.5\columnwidth]{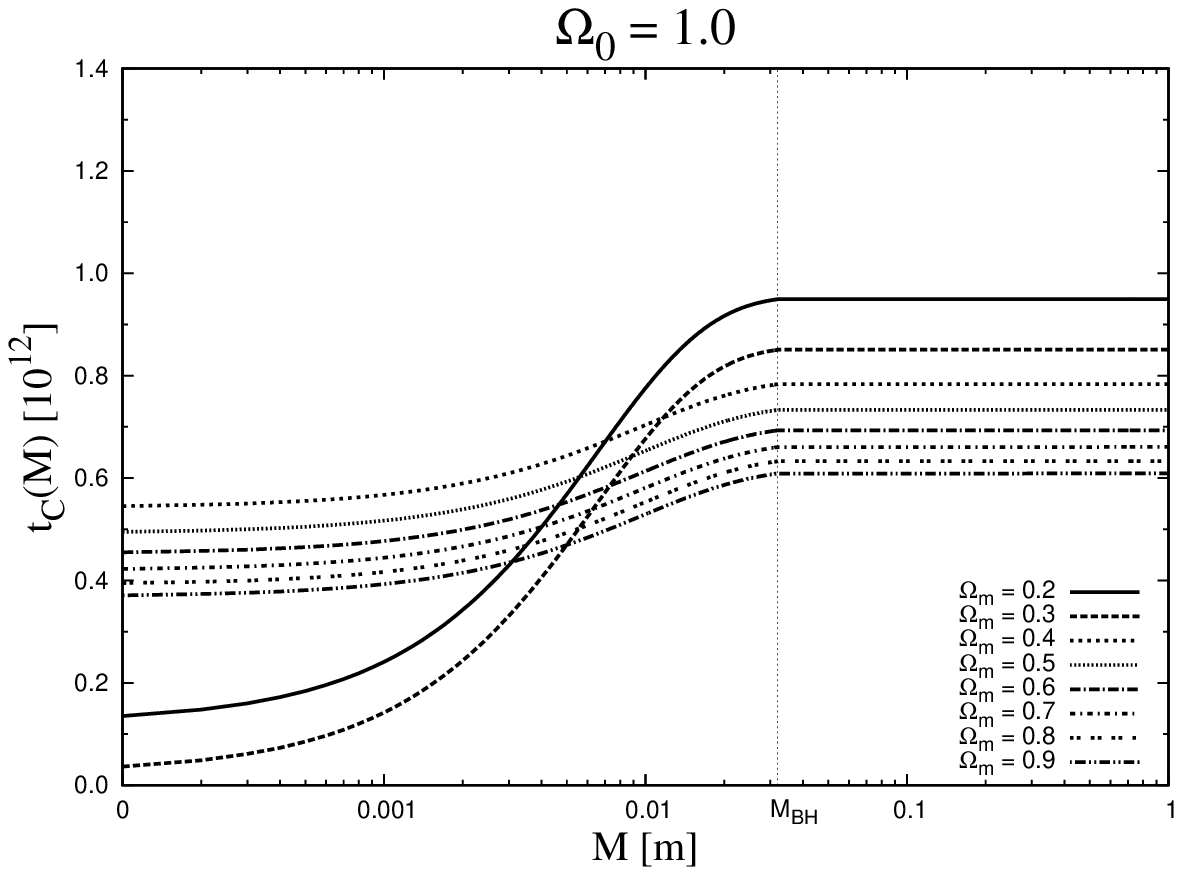}
\\
\includegraphics[width=0.5\columnwidth]{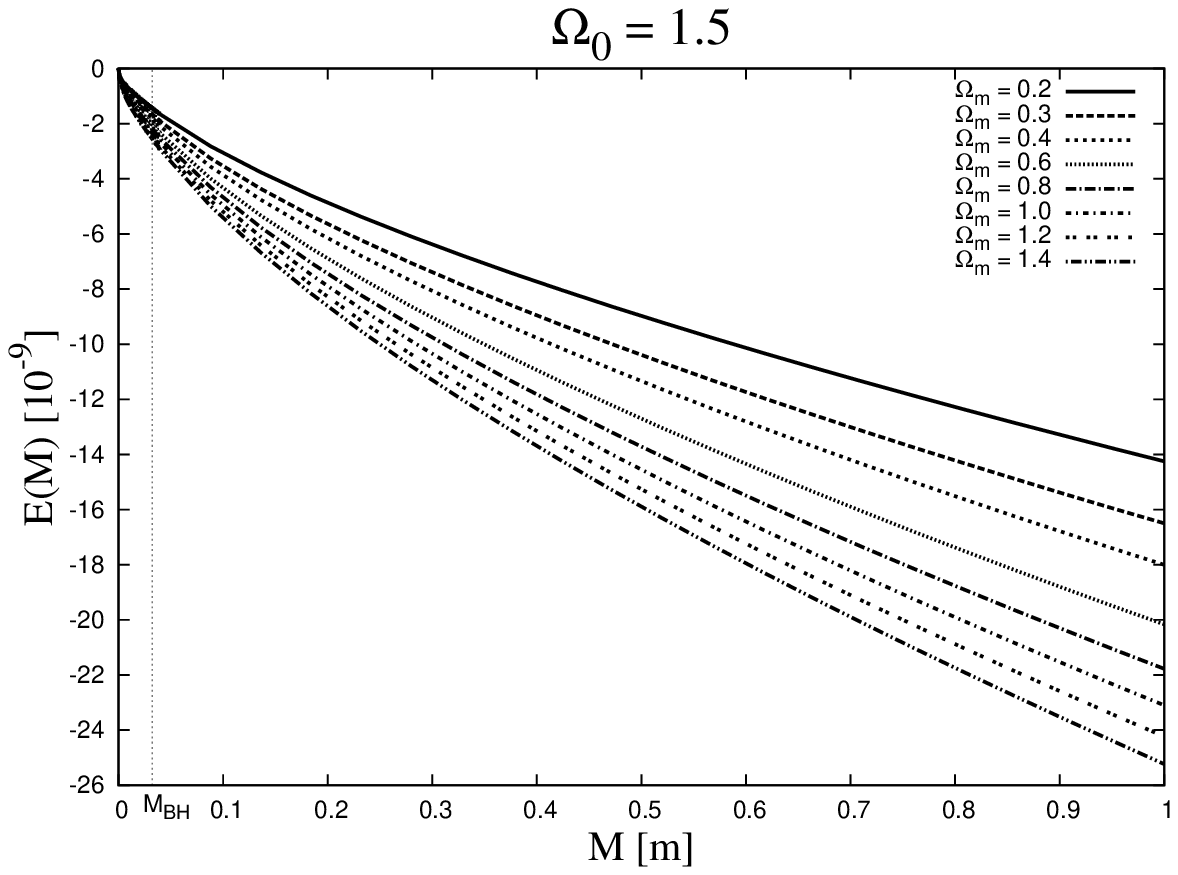}
\includegraphics[width=0.5\columnwidth]{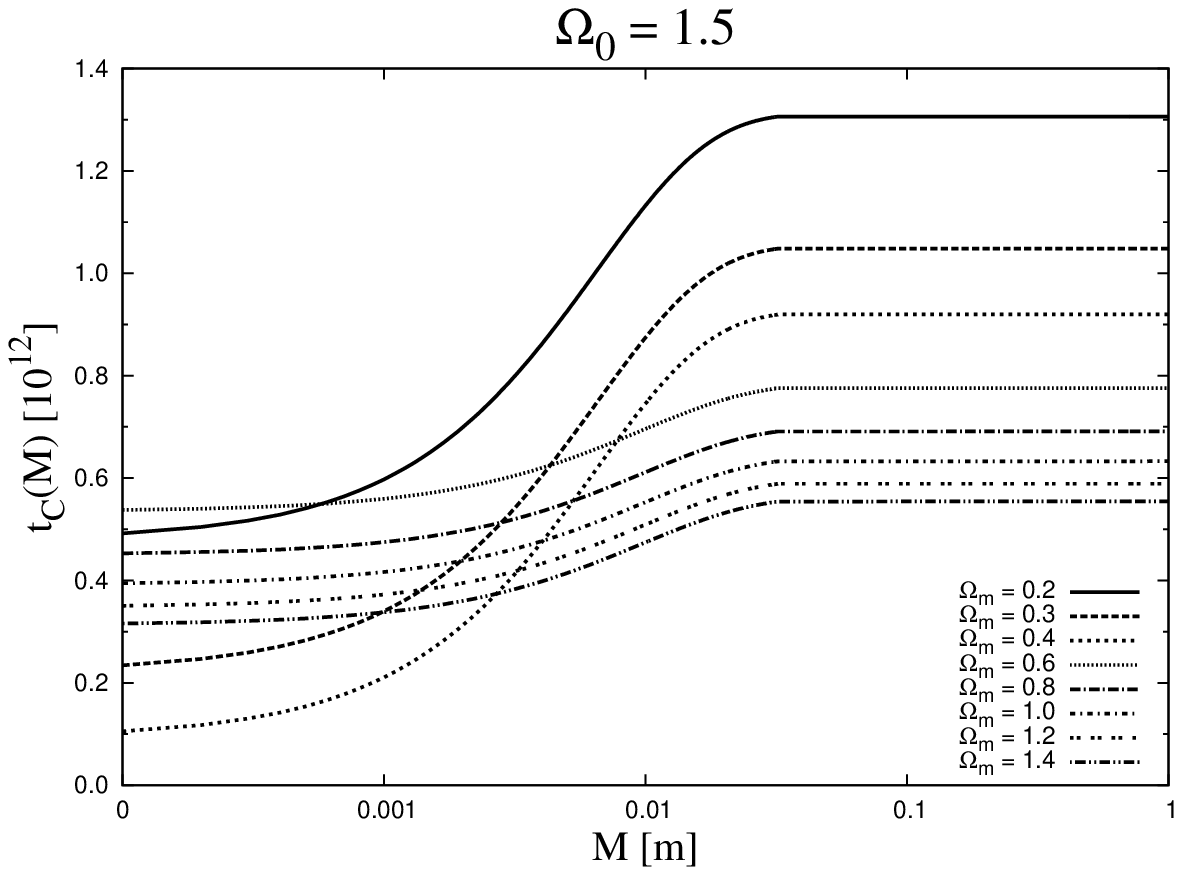}
\caption{The plot shows the $E(M)$ (left) and $t_C(M)$ (right) functions for the
same three values of $\Omega_0$ as in
Fig.~\ref{Fig:const_omega_zero_analysis-BH_age}. All the plots of $E(M)$ are
similar and do not reveal the existence of the two values of the age of the black hole. The crunch time functions show this fact by
the occurrence of the two distinctive families of curves. The curves starting at
lower values correspond to the higher value of the maximal age of the black
hole. Note the logarithmic scale on the $x$ axis. The 'flatness' of the curves
for $M>M_{\rm BH}$ is the consequence of the logarithmic $x$ axis - in reality
the curves are not flat, but change very slowly compared to the part inside the
black hole.} \label{Fig:const_omega_zero_analysis-functions}
\end{figure}
}

The results presented so far were obtained for a flat velocity and flat density
profiles at the recombination consistent with the $\Lambda$CDM model with $\Omega_m=0.266$ and $\Omega_\Lambda=0.734$. However,
the specific initial FLRW model should not play a key role in the creation of Gyrs old black holes.
Therefore, we have performed calculations of the maximal age of the black hole
for two types of evolution for initial models with different parameters. We set the Hubble parameter value at
$H_0=71$~km~s$^{-1}$~Mpc$^{-1}$ and the redshift of the CMB at $z_{\rm
rec}=1090$, based on WMPA 7--year results, and carried out the calculations for
varying $\Omega_m$ and $\Omega_\Lambda$ values.
\subsubsection{The age of the black hole}

In Fig.~\ref{Fig:const_omega_zero_analysis-BH_age} the dependence of the maximal
age of the black hole on the matter density parameter $\Omega_m$, for three
constant values of $\Omega_0$ equal to $0.5$, $1.0$ and $1.5$, is shown. For
both evolution types we can clearly see the two constant values of the maximal
age of the black hole equal to $3.717$ and $12.708$~Gyr. For $\Omega_0=0.5$ only
the first value is obtained. The shapes of the curves for the same value of
$\Omega_0$ and different evolution types are very similar. The differences are
caused by numerical errors and are remnants of the method of choosing initial
conditions for the optimization method involving random numbers. These
differences are presented in the insets in the $\Omega_0=1.5$ plots.

\subsubsection{The LT functions}

Figure~\ref{Fig:const_omega_zero_analysis-functions} shows the $E(M)$ and $t_C(M)$ functions for the same three values of $\Omega_0$ as in
Fig.~\ref{Fig:const_omega_zero_analysis-BH_age} for the evolution form flat
velocity at the recombination. For the $E(M)$ function all the curves are
similar, without any hints of the two values of the age of black holes. The plots for the crunch time function reveal the existence of the two values, as for all values of $\Omega_0$ we can indicate the two distinctive
families of curves. Because of the fact that the crunch time function and the
future apparent horizon $t_{\rm AH+}$ almost coincide in all the cases and
taking into account that the value of the future apparent horizon at its minimum
is the creation time of the black hole, we can say that the crunch time curves
starting at lower values correspond to the higher value of the maximal age of
the black hole. The reason why all the crunch time curves corresponding to one
value of the age of the black hole, e.g. $12.708$~Gyr, do not start at the same
point is that with varying $\Omega_m$ the age of the Universe changes.
\section{Summary and conclusions}

We have demonstrated that within the inhomogeneous LT model a perturbation of
the velocity or density profile at the recombination can evolve into a
spherically symmetric galaxy--like object with a black hole at the center of the
age of Gyr. This work is a refinement to the model presented in
\cite{Krasinski:2004a}, which allows for such an early creation of black holes,
together with the usage of an arbitrary density profile describing the present
day galaxy and an arbitrary FLRW model as the background matter reservoir. The
evolution of such a perturbation leading to the present mass distribution of a
galaxy progresses without any shell crossing singularities. In the model, the
mass at the center increases in consequence of different expansion rates in the central region and farther away. The crucial
point for the time of the black hole creation is the interior geometry of the
black hole, that is the bang time function $t_B$ and the crunch time function
$t_C(M)$ in the region $M<M_{\rm BH}$. By an adequate adjustment of those
arbitrary functions, we were able to find an LT model yielding the desired age
of Gyrs. It was found that the maximal age of the black hole as a function of
the density parameters $\Omega_m$ and $\Omega_\Lambda$ takes a constant value
irrespective of the $\Omega_0$ value.

The main target for the model was the proper simulation of the black hole at the
center of M87 galaxy. This supermassive black hole is one of the biggest
supermassive black holes found. We have used a new density profile describing
the mass distribution in this galaxy together with an updated value of the mass
of the black hole based on \cite{Macchetto:1997}. This required a slight
improvement of the basic model allowing to use any density profile, not only the
one for which analytical calculations are possible. We have successfully
obtained an LT model characterizing the evolution leading to this galaxy
together with the black hole created about $12.7$ Gyrs ago. This number was
independent on the type of the initial perturbation (velocity or density)
seeding the galaxy formation. By changing the parameters of the initial FLRW model we found that the maximal age of the black hole for this particular
galaxy can take only two values: $3.717$ and $12.708$. A set of thorough runs
for three values of $\Omega_0$ equal to $0.5$, $1.0$ and $1.5$, determining the
three families of FLRW universes: open, flat and closed respectively, showed
that the LT model functions behave properly in all cases.

However, as found by Gebhardt and Thomas \cite{2009ApJ...700.1690G}, the mass of
the black hole in M87 can be even $6.4~\pm~0.5~\cdot~10^9~M_\odot$, that is
twice the mass used in our calculations. This can affect the results presented
here and we plan to perform the simulations with this new value of $M_{\rm BH}$
in near future.

The main disadvantage of the LT model in simulating the evolution and creation
of galaxies is the absence of rotation. Rotation is a key factor in galaxy
formation, as it slows down the collapse and produces accretion disks around
black holes. Due to this it plays a much bigger role in the later stages of
evolution, when the collapse of the central body has started.

\bibliographystyle{spphys}
\bibliography{LT-BH}
\end{document}